\documentclass[aps,amssymb,groupedaddress,preprint,showpacs]{revtex4}
\newcommand\ket[1]{\mid#1\rangle}
\newcommand\bra[1]{\langle#1\mid}

\usepackage{dcolumn}
\usepackage{amsmath}

\usepackage{subeqnarray}
\usepackage{bm}
\begin{document}

\title{Dynamics-Controlled Truncation Scheme for Quantum Optics and Nonlinear Dynamics in Semiconductor Microcavities}
\author{S. Portolan$^{1,3,}$\footnote{Electronic address: stefano.portolan@epfl.ch}, O. Di Stefano$^2$, S. Savasta$^2$, F. Rossi$^3$, and R. Girlanda$^2$}
\affiliation{$^1$Institute of Theoretical Physics, Ecole
Polytechnique F\'{e}d\'{e}rale de Lausanne EPFL, CH-1015 Lausanne,
Switzerland} \affiliation{$^2$Dipartimento di Fisica della Materia e
Tecnologie Fisiche Avanzate, Universit\`{a} di Messina Salita
Sperone 31, I-98166 Messina, Italy} \affiliation{$^3$Dipartimento di
Fisica, Politecnico di Torino, Corso Duca degli Abruzzi 24, I-10129
Torino, Italy}

\begin{abstract}
We present a systematic theory of Coulomb-induced correlation
effects in the nonlinear optical processes within the
strong-coupling regime. In this paper we shall set a dynamics
controlled truncation scheme \cite{Axt Stahl} microscopic treatment
of nonlinear parametric processes in SMCs including the
electromagnetic field quantization. It represents the starting point
for the microscopic approach to quantum optics experiments in the
strong coupling regime without any assumption on the quantum
statistics of electronic excitations (excitons) involved. We exploit
a previous technique, used in the semiclassical context, which, once
applied to four-wave mixing in quantum wells, allowed to understand
a wide range of observed phenomena \cite{Sham PRL95}. We end up with
dynamical equations for exciton and photon operators which extend
the usual semiclassical description of Coulomb interaction effects,
in terms of a mean-field term plus a genuine non-instantaneous
four-particle correlation, to quantum optical effects.

\end{abstract}

\maketitle

\newpage
\section{Introduction}

Since the early Seventies \cite{Esaki Tsu} researchers have been
exploring the possible realization of semiconductor-based
heterostructures, devised according to the principles of quantum
mechanics. The development of sophisticated growth techniques
started a revolution in semiconductor physics, determined by the
possibility of confining electrons in practical structures. In
addition, the increasing ability in controlling fabrication
processes has enabled the manipulation of the interaction between
light and semiconductors by engineering, in addition to the
electronic wave functions, the light modes.

Entanglement is one of the key features of quantum information and
communication technology \cite{Nielsen-Chuang} and a hot topic in
quantum optics too. Parametric down-conversion is the most
frequently used method to generate highly entangled pairs of photons
for quantum-optics applications, such as quantum cryptography and
quantum teleportation. Rapid development in the field of quantum
information requires monolithic, compact sources of nonclassical
photon states enabling efficient coupling into optical fibres and
possibly electrical injection. Semiconductor-based sources of
entangled photons would therefore be advantageous for practical
quantum technologies. The strong light-matter interaction in these
systems gives rise to cavity polaritons which are hybrid
quasiparticles consisting of a superposition of cavity photons and
quantum well (QW) excitons \cite{Weisbuch-Houdre}. Demonstrations of
parametric amplification and parametric emission in semiconductor
microcavities (SMCs) with embedded QWs\cite{Baumberg, Erland,
Langbein PRB2004}, together with the possibility of ultrafast
optical manipulation and ease of integration of these microdevices,
have increased the interest on the possible realization of
nonclassical cavity-polariton states \cite{squeezing
Quattropani,CiutiBE, Savasta PRL2005,LosannaCC,SSC Savasta}. In 2004
squeezed light generation in SMCs in the strong coupling regime has
been demonstrated \cite{Giacobino}. In 2005 an experiment probing
quantum correlations of (parametrically emitted) cavity polaritons
by exploiting quantum complementarity has been proposed and realized
\cite{Savasta PRL2005}. Specifically, it has been shown that
polaritons in two distinct idler modes interfere if and only if they
share the same signal mode so that which-way information cannot be
gathered, according to Bohr's quantum complementarity principle.

Laser spectroscopy in semiconductors and in semiconductor quantum
structures has been greatly used because exciting with ultrashort
optical pulses in general results in the creation of coherent
superpositions of many-particle states. Thus it constitutes a very
promising powerful tool for the study of correlation and an ideal
arena for semiconductor cavity quantum electrodynamics (cavity QED)
experiments as well as coherent control, manipulation, creation and
measurement of non-classical states \cite{AxtKuhn,Sham PRL95,Nature
CuCl,Savasta PRL2005}. The analysis of nonclassical correlations in
semiconductors constitutes a challenging problem, where the physics
of interacting electrons must be added to quantum optics and should
include properly the effects of noise and dephasing induced by the
electron-phonon interaction and the other environment channels
\cite{Kuhn-Rossi PRB 2005}. The nonlinear optical properties of
exciton-cavity system play a key role in driving the quantum
correlations and the nonclassical optical phenomena. The crucial
role of many-particle Coulomb correlations in semiconductors marks a
profound difference from the nonlinear optics of dilute atomic
systems, where the optical response is well described by independent
transitions between atomic levels, and the nonlinear dynamics is
governed only by saturation effects mainly due to the balance of
populations between different levels.

The Dynamics Controlled Truncation Scheme (DCTS) provides a (widely
adopted) starting point for the microscopic theory of the
light-matter interaction effects beyond mean-field \cite{AxtKuhn},
supplying a consistent and precise way to stop the infinite
hierarchy of higher-order correlations which always appears in the
microscopic approaches of many-body interacting systems without need
to resort to any assumption on the quantum statistics of the
quasi-particle arising in due course. By exploting this scheme,  it
was possible to express nonlinearities originating from the Coulomb
interaction as an instantaneous mean-field exciton-exciton
interaction plus a  noninstantaneous term where four-particle
correlation effects beyond menfield are contained entirely in a
retarded memory function \cite{Sham PRL95}. In 1996 the DCTS was
extended in order to include in the description the quantization of
the electromagnetic field and polariton effects \cite{Savasta
PRL96}. This extension has been applied to the study of quantum
optical phenomena in semiconductors and it was exploited to predict
polariton entanglement \cite{SSC Savasta}. The obtained equations
showed that quantum optical correlations (as  nonlinear optical
effects) arise from both saturation effects (phase-space filling)
and Coulomb induced correlations due to four-particle states
(including both bound and unbound biexciton states). The dynamical
equations included explicitly biexciton states. The structure of
those equations didn't allow the useful separation of Coulomb
interaction in terms of a mean-field interaction term plus a
noninstantaneous correlation term performed in the semiclassical
description.

In this paper we shall set a DCTS microscopic treatment of nonlinear
parametric processes in SMCs including the light-field quantization.
It represents the starting point for the microscopic approach to
quantum optics experiments in the strong coupling regime. For this
purpose we shall exploit a previous technique \cite{Sham PRL95}
which, once applied to four-wave mixing in QWs, it allowed to
understand a wide range of observed phenomena. Here all the
ingredients contributing to the dynamics are introduced and
commented. We shall give in great details the manipulations required
in order to provide an effective description of the nonlinear
parametric contributions beyond mean-field in an exciton-exciton
correlation fashion. In particular we derive the coupled equations
of motion for the excitonic polarization and the intracavity field.
It shows a close analogy to the corresponding equation describing
the semiclassical (quantized electron system, classical light field)
coherent $\chi^{(3)}$ response in a QW \cite{Sham PRL95}, the main
difference being that here the (intracavity) light field is regarded
not as a driving external source but as a dynamical field
\cite{Savasta PRL2003}. This correspondence is a consequence of the
linearization of quantum fluctuations in the nonlinear source term
here adopted, namely the standard linearization procedure of quantum
correlations adopted for large systems \cite{Walls}. However the
present approach includes the light field quantization and can thus
be applied to the description of quantum optical phenomena. Indeed,
striking differences between the semiclassical and the full quantum
descriptions emerge when considering expectation values of exciton
and photon numbers or even higher order correlators, key quantities
for the investigation of coherence properties of quantum light
\cite{Savasta PRL2005}. This is the main motivation for the
derivation of fully operatorial dynamical equations, within such
lowest order nonlinear coherent response, we address in the last
section. The results here presented provide a microscopic
theoretical starting point for the description of quantum optical
effects in interacting electron systems with the great accuracy
accomplished for the description of the nonlinear optical response
in such many-body systems, see e.g. \cite{Sham PRL95,Savasta
PRL2003,Savasta PRB2001,Buck,AxtKuhn} and references therein. The
proper inclusion of the detrimental environmental interaction, an
important and compelling issue, is left for a detailed analysis in
another paper of ours \cite{nostro PRB}.

In Section \ref{1} the generality of the coupled system taken into
account are exposed, here all the ingredients contributing to the
dynamics are introduced and commented. The linear and the lowest
nonlinear dynamics is the subject of Sec. \ref{2}, whereas in Sec.\,
\ref{3} we shall give in great details the manipulations required in
order to provide an effective description of the nonlinear
parametric contributions beyond mean-field in an exciton-exciton
correlation fashion. In Sec. \ref{4} the operatorial equations of
motion for exciton and intracavity photon operators are derived.
%
%
%
%
%
%================================================
\section{The Coupled System}\label{1}
%================================================
%
%
%
The system we have in mind is a semiconductor QW grown inside a
semiconductor planar Fabry-Perot resonator. In the following we
consider a zinc-blende-like semiconductor  band structure. The
valence band is made from $p$-like ($l=1$) orbital states which,
after spin-orbit coupling, give rise to $j = 3/2$ and $j=1/2$
decoupled states. In materials like GaAs, the upper valence band is
fourfold degenerate ($j=3/2$), whereas in GaAs-based QWs the valence
subbands with $j=3/2$ are energy splitted into two-fold degenerate
heavy valence subbands  with $j_z=\pm 3/2$ and light lower energy
subbands with $j_z= \pm 1/2$. The conduction band, arising from an
$s$-like orbital state (l=0), gives rise  to $j=1/2$ twofold states.
In the following we will consider for the sake of simplicity only
twofold states from the upper valence  and lowest conduction
subbands. As a consequence electrons in a conduction band as well as
holes have an additional spin-like degree of freedom as electrons in
free space. When necessary both heavy and light hole valence bands
or subbands can be included in the present semiconductor model. Only
electron-hole ({\em{eh}}) pairs with total projection of angular
momentum $\sigma = \pm 1$ are dipole active in optical interband
transitions.  In GaAs QWs photons with  circular polarizations
$\sigma = -$($+$) excite electrons with $j_z^{\it e}=+1/2$
($j_z^{\it e}=-1/2$) and holes with $j_z^{\it h}=-3/2$ ($j_z^{\it
h}=3/2$). We  label optically active {\em{eh}} pairs with the same
polarization label of light generating them; e.g. $\sigma = +1$
indicates an {\em{eh}} pair with $j_z^{\it e}=-1/2$ and $j_z^{\it
h}=3/2$.

We start from the usual model for the electronic Hamiltonian of
semiconductors \cite{Haugh,AxtKuhn}. It is obtained from the
many-body Hamiltonian of the interacting electron system in a
lattice, keeping explicitly only those terms in the Coulomb
interaction preserving the number of electrons in a given band, see
Appendix \ref{Npair states}. The system Hamiltonian can be rewritten
as
\begin{equation}\label{Ham electron} \hat{H}_e = \hat{H}_0 +
\hat{V}_{\text{Coul}}= \sum_{N \alpha} E_{N \alpha} \ket{{N \alpha}}
\bra{{N \alpha}}\, ,
\end{equation} where the eigenstates of $\hat{H}_e$, with energies
$E_{N \alpha} =\hbar \omega_{N \alpha}$, have been labelled
according to the number N of {\em eh} pairs. The state $\ket{{N=0}}$
is the electronic ground state, the $N=1$ subspace is the exciton
subspace with the additional collective quantum number $\alpha$
denoting the exciton energy level $n$, the in-plane wave vector
${\bf k}$ and the spin index $\sigma$. When needed we will adopt the
following notation: $\alpha\equiv (n,k)$ with $k\equiv ({\bf k},
\sigma)$. In QWs, light and heavy holes in valence band are split
off in energy. Assuming that this splitting is much larger than
kinetic energies of all the involved particles and, as well, much
larger than the interaction between them, we shall consider only
heavy hole states as occupied. On the contrary to the bulk case, in
a QW single particle states experience confinement along the growth
direction and subbands appear, anyway in the other two orthogonal
directions translational invariance is preserved and the in-plane
exciton wave vector remains a good quantum number. Typically, the
energy difference between the lowest QW subband level and the first
excited one is larger than the Coulomb interaction between
particles, and  we will consider excitonic states arising from
electrons and heavy holes in the lowest subbands.

Eigenstates of the model Hamiltonian with N=1 (called excitons) can
be created from the ground state by applying the exciton creation
operator:
\begin{equation}\label{exciton def} \bigl| 1 n \sigma {\bf k}\bigr>=\hat{B}^\dagger_{n \sigma
{\bf k}} \bigl| N=0\bigr>\, ,\end{equation}
which can be written in terms of electrons and holes  operators as
\begin{equation} \hat{B}^\dagger_{n
\sigma {\bf k}} = \sum_{{\bf k}'} \Phi^{\bf k}_{n \sigma {\bf k}'}
\hat{c}^\dagger_{\sigma, {\bf k}' + \eta_e{\bf
k}/2}\hat{d}^\dagger_{\sigma, -{\bf k}' + \eta_h{\bf k}/2}\, ,
\label{Bdag}\end{equation}
here $\Phi^{\bf k}_{n \sigma {\bf k}'}$ is the exciton wave
function, being ${\bf k}$ the total wave vector ${\bf k} = {\bf k}_e
+ {\bf k}_h$, and ${\bf k}' = \eta_e{\bf k}_e - \eta_h{\bf k}_h$
with $\eta_{(e,h)} = m_{(e,h)}/(m_{(e)} + m_{(h)})$ ($m_{e}$ and
$m_{h}$ are the electron and hole effective masses). These exciton
eigenstates can be obtained by requiring the general one {\em eh}
pair states to be eigenstates of $\hat{H}_e$:
\begin{equation}
\hat{H}_e \bigl| 1 n \sigma {\bf k}\bigr> = \hbar \omega_{1 n \sigma
{\bf k}} \bigl| 1 n \sigma {\bf k}\bigr>\, ,
\label{secular}\end{equation} and projecting this secular equation
onto the set of product ({\em eh}) states $\bigl|k_e,k_h\bigr> =
\hat{c}^\dag_{k_e} \hat{d}^\dag_{k_h} \ket{0}$, (see Appendix
\ref{Npair states} for details):
\begin{equation}\label{Schrodinger} \sum_{k_e,{k}_h} ( \bra{k'_e,k'_h}\hat{H}_c \ket{k_e
,k_h} - \hbar \omega_{n \sigma {\bf k}} \delta_{k'_e k'_h,k_e k_h} )
\bigl< k_e,k_h \bigl| 1 n \sigma {\bf k}\bigr>  = 0\, .
\end{equation} Thus, having expressed the correlated exciton state as a
superposition of uncorrelated product states,
\begin{equation} \bigl| 1 n \sigma {\bf k}\bigr> = \sum_{k_e,k_h} {\Bigg (}\bigl< k_e,k_h \bigl| 1 n \sigma{\bf
k}\bigr> {\Bigg )}\bigl|k_e,k_h\bigr> \, ,\end{equation}
the scalar products, coefficients of this expansion, represent
nothing but the envelope function $\Phi^{\bf k}_{n, \sigma, {\bf
k}'}$ of the excitonic aggregate being the solution of the
corresponding Schr\"{o}dinger equation (\ref{Schrodinger}). It
describes the correlated {\em eh} relative motion in k-space. In
order to simplify a bit the notation, the spin convention in Eq.\,
(\ref{Bdag}) has been changed by using the same label for the
exciton spin quantum number and for the spin projections of the
electron and hole states forming the exciton.

The next relevant subspace ($N=2$) is the biexciton one, spanning
all the states with 2 {\em eh} pairs. It seems worth noting that the
above description of {\em eh} complexes arises from the properties
of quantum states and, once fixed the system Hamiltonian, no
approximations have been introduced insofar. Indeed such a property
hold for any N {\em eh} pairs aggregate and we will give a full
account of it in Appendix \ref{Npair states}.

The eigenstates of the Hamiltonian $\hat{H}_{c}$ of the cavity modes
can be written as $\ket{n, \lambda}$ where $n$ stands for the total
number of photons in the state and $\lambda = ({\bf k}_1, \sigma_1;
...  ;{\bf k}_n, \sigma_n)$ specifies wave vector and polarization
$\sigma$ of each photon. Here we shall neglect the longitudinal-
transverse splitting of polaritons \cite{Kavokin} originating mainly
from the corresponding splitting of cavity modes. It is more
relevant at quite high in-plane wave vectors  and often it results
to be smaller than the polariton linewidths. The present description
can be easily extended to include it. We shall treat the cavity
field in the quasi-mode approximation, that is to say we shall
quantize the field as the mirrors were perfect and subsequently we
shall couple the cavity with a statistical reservoir of a continuum
of external modes. This coupling is able to provide the cavity
losses as well as the feeding of the coherent external impinging
pump beam. The cavity mode Hamiltonian, thus, reads
\begin{equation}\label{Ham cavity} \hat{H}_{c} = \sum_k \hbar \omega^c_k
 \hat{a}_k ^{\dag} \hat{a}_k\, , \end{equation} with the operator
$\hat{a}^\dag_k$ which creates a photon state with energy $\hbar
\omega^c_k =\hbar (\omega^2_{\text{exc}} + v^2 |{\bf k}|^2)^{1/2}$,
$v$ being the velocity of light inside the cavity and $k =
(\sigma,{\bf k})$. The coupling between the electron system and the
cavity modes is given in the usual rotating wave approximation
\cite{Savasta PRL96,HRS Savasta}
\begin{equation}\label{Ham inter cav-exc} \hat{H}_{I} = - \sum_{n k} V^*_{n k} \hat{a}_k
^{\dag} \hat{B}_{n k} + H.c.\, , \end{equation} $V_{n,k }$ is the
photon-exciton coupling coefficient enhanced by the presence of the
cavity \cite{Savona Quattropani SSC} set as $V_{n,k} = \tilde
V_{\sigma} \sqrt{A} \phi^*_{n,\sigma}({\bf x}=0)$, the latter being
the real-space exciton envelope function calculated in the origin
whereas $A$ is the in-plane quantization surface, $\tilde
V_{\sigma}$ is proportional to the interband dipole matrix element.
Modeling the loss through the cavity mirrors within the quasi-mode
picture means we are dealing with an ensemble of external modes,
generally without a particular phase relation among themselves. An
input light beam impinging on one of the two cavity mirrors is an
external field as well and it must belong to the family of modes of
the corresponding side (i.e. left or right). Being coherent, it will
be the non zero expectation value of the ensemble. It can be shown
\cite{Savasta PRL96,nostro PRB} that for a coherent input beam, the
driving of the cavity modes may be described by the model
Hamiltonian \cite{Savasta PRL96,nostro PRB}
\begin{equation}\label{H quasi modi} \hat{H}_p = i\, t_c \sum_{\bf k} ({E}_{\bf k} \hat{a}^\dag_{\bf k} -
{E}^{*}_{\bf k} \hat{a}_{\bf k})\, ,\end{equation} where ${E}_{\bf
k}$ (${E}^*_{\bf k}$) is a $\mathbb{C}$-number describing the
positive (negative) frequency part of the coherent input light field
amplitude.

\section{Linear and Nonlinear Dynamics}\label{2}

The idea is not to use a density matrix approach, but to derive
directly expectation values of all the quantities at play. The
dynamics is described by ``transition" operators (known as
generalized Hubbard operators):
\begin{eqnarray}\label{Hubbard} \hat{X}_{N,\alpha;M,\beta} =
\ket{{N,\alpha}} \bra{{M,\beta}} \nonumber \\
\hat{Y}_{n,\lambda;m,\mu}= \ket{n,\lambda} \bra{m,\mu}\, .
\end{eqnarray} The fundamental point in the whole analysis is that,
thanks to the form of the interaction Hamiltonian $\hat{H}_I$ and
thanks to the quasiparticle conservation the free Hamiltonians
possess, we can use the so-called dynamics controlled truncation
scheme, stating that we are facing a rather special model where the
correlation have their origin only in the action of the
electromagnetic field and thus the general theorem due to Axt and
Stahl \cite{Axt Stahl} holds. For our purpose we will need its
generalization in order to include the quantization of the
electromagnetic field \cite{Savasta PRL96}, it reads:
\begin{eqnarray}\label{DCTS} \langle \hat{X}_{N,\alpha;M,\beta}
\hat{Y}_{n,\lambda;m,\mu} \rangle = \sum_{i=0}^{i_0} \langle
\hat{X}_{N,\alpha;M,\beta} \hat{Y}_{n,\lambda;m,\mu} \rangle
^{(N+M+n+m+2i)} \nonumber \\+ \mathcal{O}(E^{(N+M+n+m+2i_0+2)})\, ,
\end{eqnarray} i.e. the expectation value of a \textbf{zero to N-pair}
transition is at least of order N in the external electromagnetic
field. There are only even powers because of the spatial inversion
symmetry which is present. Once a perturbative order in the external
coherent fields is chosen, Eq.\, (\ref{DCTS}) limits the expectation
values to take into account, thus providing a precise way to
truncate the hierarchy of equations of motions.

The exciton and photon operators can be expressed as
\begin{eqnarray}\label{a,B} \hat{a}_k &=& \hat{Y}_{0;1 k} + \sum_{n \geq 1}
\sqrt{n_k+1} \hat{Y}_{n_k k; (n_k+1) k} \nonumber \\
\hat{B}_{n k} &=& \hat{X}_{0;1 n k} + \sum_{N \geq 1, \alpha \beta}
\bra{N \alpha}\hat{B}_{n k} \ket{(N+1) \beta} \hat{X}_{N
\alpha;(N+1) \beta}\, , \end{eqnarray} where in writing the photon
expansion we omitted all the states not belonging to the $k$-th mode
which add up giving the identity in every Fock sector \cite{detail}.

The equation of motion for the generic quantity of interest
$\hat{X}_{N,\alpha;M,\beta} \hat{Y}_{n,\lambda;m,\mu}$ is reported
in Appendix \ref{gen eq}. In the Heisenberg picture we start
considering the equation of motion for the photon and exciton
operators, once taken the expectation values we exploit theorem
(\ref{DCTS}) retaining only the linear terms. With the help of the
generalized Hubbard opertors all this procedure may be done by
inspection. The linear dynamics for $ \left<\right. \hat{a}_k
\left.\right>^{(1)} = \left<\right. \hat{Y}_{0;1 n k}
\left.\right>^{(1)}$ and $ \left<\right. \hat{B}_{n k}
\left.\right>^{(1)} = \left<\right. \hat{X}_{0;1 n k}
\left.\right>^{(1)}$ reads:
\begin{eqnarray}\label{lin order}
&& \frac{d}{dt}\left<\right. \hat{a}_k \left.\right>^{(1)} = -i
\bar{\omega}^c_k \left<\right. \hat{a}_k \left.\right>^{(1)} +i
\sum_{n} \frac{V^*_{n k}}{\hbar} \left<\right. \hat{B}_{n k}
\left.\right>^{(1)} + t_c \frac{E_k}{\hbar} \\
&& \frac{d}{dt}\left<\right. \hat{B}_{n k} \left.\right>^{(1)} = -i
\bar{\omega}_{1 n k} \left<\right. \hat{B}_{n k} \left.\right>^{(1)}
+ i \frac{V_{n k}}{\hbar} \left<\right. \hat{a}_{k}
\left.\right>^{(1)} \, . \end{eqnarray}
In these equations $\bar{\omega}^c_k = \omega^c_k-i\gamma_k$, where
$\gamma_k$ is the cavity damping, analogously $\bar{\omega}_{1 n k}
= \omega_{1 n k} -i\Gamma_{\text{x}}$ and $\bar{\omega}_{2 \beta} =
\omega_{2 \beta}-i\Gamma_{\text{xx}}$. The dynamics up to the third
order is a little bit more complex, we shall make extensively use of
(\ref{dt gen op}) (in the following the suffix $^{+(n)}$ stands for
``up to" $n$-th order terms in the external electromagnetic exciting
field). With Eq.\, (\ref{a,B}) the exciton and the photon
expectation values can be expanded as follows:
\begin{equation}\label{B up to 3}
\left<\right. \hat{B}_{n k} \left.\right>^{+(3)} = \left<\right.
\hat{X}_{0;1 n k} \left.\right>^{+(3)} + \sum_{\alpha \beta}\bra{1
\alpha} \hat{B}_{n k} \ket{2 \beta} \left<\right. \hat{X}_{1
\alpha;2 \beta} \left.\right>^{(3)}\, , \end{equation}
\begin{equation}\label{a up to 3}
\left<\right. \hat{a}_{k} \left.\right>^{+(3)} = \left<\right.
\hat{Y}_{0;1 k} \left.\right>^{+(3)} + \sqrt{2} \left<\right.
\hat{Y}_{1 k;2 k} \left.\right>^{(3)}\, .
\end{equation}
With a bit of algebra we obtain
\begin{equation}\label{dt a up to 3}
\frac{d}{dt}\left<\right. \hat{a}_k \left.\right>^{+(3)} = -i
\bar{\omega}^c_k \left<\right. \hat{a}_k \left.\right>^{+(3)} + i
\sum_{n} \frac{V^*_{n k}}{\hbar} \left<\right. \hat{B}_{n k}
\left.\right>^{+(3)} + t_c \frac{E_k}{\hbar}\, , \end{equation}
\begin{eqnarray}\label{dt B up to 3}
&&\frac{d}{dt}\left<\right. \hat{B}_{n k} \left.\right>^{+(3)} = -i
\bar{\omega}_{1 n k} \left<\right. \hat{B}_{n k}
\left.\right>^{+(3)} + i \frac{V_{n k}}{\hbar} \left<\right.
\hat{a}_{k} \left.\right>^{+(3)} + \nonumber \\
&& \hspace{2.0cm} + \sum_{\tilde{n} \tilde{k}} {\Bigg [}
\frac{i}{\hbar} \sum_{n' k', \alpha} V_{n' k'} \bra{1 \tilde{n}
\tilde{k}} [\hat{B}_{n k}, \hat{B}^\dag_{n' k'} ] - \delta_{(n'
k');(n k)} \ket{1 \alpha} \langle \hat{X}_{1 \tilde{n} \tilde{k}; 1
\alpha} \hat{Y}_{0; 1 k'} \left.\right>^{(3)} - \nonumber \\
&& \hspace{3.0cm} - i \sum_{\beta}(\omega_{2 \beta} - \omega_{1
\tilde n \tilde k} - \omega_{1 n k}) \bra{1 \tilde n \tilde k}
\hat{B}_{n k} \ket{2 \beta} \langle \hat{X}_{1 \tilde n \tilde k; 2
\beta} \hat{Y}_{0;0} \left.\right>^{(3)} {\Bigg ]}\, ,\end{eqnarray}
in analogy with the eqs \cite{Savasta PRL96} (see also Ref.\,
\cite{Sham PRL95}). The resulting equation of motion for the lowest
order biexciton amplitude is
\begin{eqnarray}\label{X02}
&& \frac{d}{dt}\left<\right. \hat{X}_{0;2 \beta} \left.\right>^{(2)}
= -i \bar{\omega}_{2 \beta} \left<\right.
\hat{X}_{0;2 \beta} \left.\right>^{(2)} + \nonumber \\
&& \hspace{3.0cm} + \frac{i}{\hbar} \sum_{n' k';n'' k''} V_{n'
k'}\bra{2 \beta} \hat{B}^\dag_{n' k'} \ket{1 n'' k''} \left<\right.
\hat{X}_{0,1 n'' k''} \hat{Y}_{0,1 k'} \left.\right>^{(2)}\,
.\end{eqnarray}

\section{Coherent Response}\label{3}

Thanks to the fact we want to treat coherent optical processes it is
possible to manipulate further the parametric contributions under
two assumptions. We are addressing a coherent optical response, thus
we may consider that a coherent pumping mainly generates
\textit{coherent} nonlinear processes, as a consequence the dominant
contribution of the biexciton sector on the third-order nonlinear
response can be calculated considering the system quantum state as a
pure state, which means the nonlinear term is regarded as
originating mainly from coherent contributions. Moreover
nonclassical correlations are taken into account up to the lowest
order. The first assumption results in the factorizations $\langle
\hat{X}_{1 \tilde n \tilde k; 2 \beta} \hat{Y}_{0;0}
\left.\right>^{(3)} \simeq \langle \hat{X}_{1 \tilde n \tilde k; 0}
\rangle^{(1)} \langle \hat{X}_{0; 2 \beta} \left.\right>^{(2)}$ and
$\langle \hat{X}_{1 \tilde{n} \tilde{k}; 1 \beta} \hat{Y}_{0; 1 k'}
\left.\right>^{(3)} \simeq \langle \hat{X}_{1 \tilde{n}
\tilde{k};0}\rangle^{(1)} \langle \hat{X}_{0; 1 \beta} \hat{Y}_{0; 1
k'} \left.\right>^{(2)}$. The second implies $\langle \hat{X}_{0; 1
\beta} \hat{Y}_{0; 1 k'} \left.\right>^{(2)} \simeq \langle
\hat{X}_{0;1 \beta}\rangle^{(1)} \langle \hat{Y}_{0; 1 k'}
\left.\right>^{(1)}$, in the nonlinear source term, namely the
standard linearization procedure of quantum correlations adopted for
large systems \cite{Walls}. Of course these two approximations can
be avoided at the cost of enlarging the set of coupled equations in
order to include the equation of motions for the resulting
correlation functions. It neglects higher order quantum optical
correlation effects between the electron system and the cavity modes
leading to a renormalization of the biexciton dynamics with
intriguing physical perspectives. However for extended systems, like
QWs in planar microcavities, these are effects in most cases of
negligible impact, on the contrary in fully confined geometries such
as cavity embedded quantum dots they could give significant
contributions. In the end, within such a \textit{coherent limit}, we
are able to describe the biexciton contribution effectively as an
exciton-exciton correlation \cite{Sham PRL95}. The resulting
equations for the coupled exciton an cavity-field expectation values
coincide with those obtained within a semiclassical theory
(quantized electron-system and classical cavity field). Nevertheless
completely different results can be obtained for exciton or photon
number expectation values or for higher order correlation function
\cite{SSC Savasta,HRS Savasta}. In the next section we will derive
operator equations useful for the calculation of such correlation
functions. After the two approximations described above
(linearization of quantum fluctuations and coherent limit), Eqs\,
(\ref{dt B up to 3}) becomes
\begin{equation}\label{dt B up to 3fin}
\frac{d}{dt}\left<\right. \hat{B}_{n k} \left.\right>^{+(3)} = -i
\bar{\omega}_{1 n k} \left<\right. \hat{B}_{n k}
\left.\right>^{+(3)} +i \frac{V_{n k}}{\hbar} \left<\right.
\hat{a}_{k} \left.\right>^{+(3)} -\frac{i}{\hbar} \sum_{\tilde{n}
\tilde{k}} \langle \hat{B}_{\tilde n \tilde k} \rangle^{*(1)}
R^{(2)}_{n k;\tilde n \tilde k}\, ,\end{equation}
where
\begin{subeqnarray}\label{R nl}
&&R^{(2)}_{n k;\tilde n \tilde k} = Q^{\text{PSF}(2)}_{n k;\tilde n
\tilde k} + Q^{\text{COUL}(2)}_{n k;\tilde n \tilde k} \\
&&Q^{\text{PSF}(2)}_{n k;\tilde n \tilde k} = \sum_{n' k', n'' k''}
C^{n' k',n'' k''}_{\tilde n \tilde k,n k} \langle \hat{B}_{n'' k''}
\rangle^{(1)} \langle \hat{a}_{k'} \rangle^{(1)}\\
&&Q^{\text{COUL}(2)}_{n k;\tilde n \tilde k} =
\sum_{\beta}c^{(1)}_{n k;\tilde n \tilde k;\beta} \langle
\hat{X}_{0; 2 \beta} \rangle^{(2)}\, ,
\end{subeqnarray} with
\begin{eqnarray}\label{coeffs}
&& C^{n' k',n'' k''}_{\tilde n \tilde k,n k} = V_{n' k'} \bra{1
\tilde{n} \tilde{k}} \delta_{(n' k');(n k)} - [\hat{B}_{n k}, \hat{B}^\dag_{n' k'} ] \ket{1 n'' k''} \\
&& c^{(1)}_{n k;\tilde n \tilde k;\beta} = \hbar (\omega_{2 \beta} -
\omega_{1 \tilde n \tilde k} - \omega_{1 n k}) \bra{1 \tilde n
\tilde k} \hat{B}_{n k} \ket{2 \beta}\, .
\end{eqnarray}
This equation is analogous to the corresponding equation describing
the semiclassical (quantized electron system, classical light field)
coherent $\chi^{(3)}$ response in a QW \cite{Sham PRL95}, the main
difference being  that here the (intracavity) light field is
regarded not as a driving external source but as a dynamical field
\cite{Savasta PRL2003}. This close correspondence for the dynamics
of expectation values of the exciton operators, is a consequence of
the linearization of quantum fluctuations. However the present
approach includes the light field quantization and can thus be
applied to the description of quantum optical phenomena.
%Indeed,
%striking difference between the semiclassical and the full quantum
%descriptions emerge when considering expectation values of exciton
%and photon numbers or even higher order correlators. For this
%reason, in the following section we derive operatorial dynamical
%equations useful for the calculation of such higher order
%correlators.
By explicit calculation it is easy to see that the first term in
Eq.\,  (\ref{coeffs}) is zero unless all the involved polarization
labels $\sigma$ coincide. In order to manipulate the last term we
follow the procedure of Ref.\, \cite{Sham PRL95} which succeeded in
reformulating the nonlinear term coming from the Coulomb interaction
as an exciton-exciton (X-X) mean-field contribution plus a
correlation term driven by a two-exciton correlation function. Even
if we are about to perform more or less the same steps of Ref.\,
\cite{Sham PRL95} we shall provide a detailed account  of all the
key points of the present derivation. A clear comprehension of these
details will be essential for the extension to operatorial dynamical
equations of the next section.

In performing this we shall need the two identities:
\begin{eqnarray}
&& c^{(1)}_{n k;\tilde n \tilde k;\beta} = \hbar (\omega_{2 \beta} -
\omega_{1 \tilde n \tilde k} - \omega_{1 n k})
\bra{1 \tilde n \tilde k} \hat{B}_{n k} \ket{2 \beta} = \nonumber \\
&& \hspace{2.5cm} = \hbar \bra{1 \tilde n \tilde k} \hat{B}_{n k}
\big( \frac{\hat{H}_{c}}{\hbar} - \omega_{1 \tilde n \tilde k} -
\omega_{1 n k} \big) \ket{2 \beta} \end{eqnarray}
and
\begin{eqnarray}\label{Sham identity}
&& \frac{d}{dt} \Bigg( \langle \hat{B}_{n' k'} \left.\right>^{(1)}
\langle \hat{B}_{n'' k''} \left.\right>^{(1)} e^{-i \Omega(u-t)}
\bigg) =  \\
&& \hspace{2.5cm} = +\frac{i}{\hbar} \bigg( V_{n' k'} \langle
\hat{a}_{k'} \rangle^{(1)} \langle \hat{B}_{n'' k''} \rangle^{(1)} +
V_{n'' k''} \langle \hat{a}_{k''} \rangle^{(1)} \langle \hat{B}_{n'
k'} \rangle \bigg)e^{-i \Omega(t-t')} \nonumber\, , \end{eqnarray}
or
\begin{equation}\label{Sham identity2}
\frac{1}{2}\ \frac{d}{dt} \sum_{n' k';n'' k''} \Bigg( \langle
\hat{B}_{n' k'} \left.\right>^{(1)} \langle \hat{B}_{n'' k''}
\left.\right>^{(1)} e^{-i \Omega(t-t')} \bigg) = +\frac{i}{\hbar}
\sum_{n' k';n'' k''} V_{n' k'} \langle \hat{a}_{k'} \rangle^{(1)}
\langle \hat{B}_{n'' k''} \rangle^{(1)} e^{-i \Omega(t-t')}\,
,\end{equation} where $\Omega \doteq \omega_{1 n' k'} + \omega_{1
n'' k''} - 2 i \Gamma_{\text{x}}\, .$
Employing the forma solution of the biexciton amplitude Eq.\,
(\ref{X02}) we have:
\begin{eqnarray}\label{conti c}
&& \hspace{-1.0cm} \sum_{\beta}c^{(1)}_{n k;\tilde n \tilde k;\beta}
\langle \hat{X}_{0; 2 \beta} \rangle^{(2)} = \hbar \sum_{\beta}
\bra{1 \tilde n \tilde k} \hat{B}_{n k} \big(
\frac{\hat{H}_{c}}{\hbar} -
\omega_{1 \tilde n \tilde k} - \omega_{1 n k} \big) \ket{2 \beta} \cdot \\
&& \hspace{0.0cm} i \sum_{n' k';n'' k''} \frac{V_{n' k'}}{\hbar}
\bra{2 \beta} \hat{B}^\dag_{n' k'} \ket{1 n'' k''} \int_{-\infty}^t
dt' e^{-i (\omega_{2 \beta} - i \Gamma_{\text{xx}})(t-t')} \langle
\hat{a}_{k'} \rangle^{(1)}(t') \langle \hat{B}_{n'' k''}
\rangle^{(1)}(t') \nonumber\, .
\end{eqnarray}
We observe that the matrix elements entering the nonlinear source
terms are largely independent on the wave vectors for the range of
wave vectors  of interest in the optical response. Neglecting such
dependence we can thus exploit the identity (\ref{Sham identity2}),
obtaining
\begin{eqnarray}\label{conti c 2}
&& \hspace{-1.0cm} = \hbar \sum_{\beta} \bra{1 \tilde n \tilde k}
\hat{B}_{n k} \big( \frac{\hat{H}_{c}}{\hbar} - \omega_{1 \tilde n
\tilde k} - \omega_{1 n k} \big) \ket{2 \beta}
\int_{-\infty}^t dt' e^{-i (\omega_{2 \beta} - i \Gamma_{\text{xx}})(t-t')} \nonumber \\
&& \hspace{0.0cm} \sum_{n' k';n'' k''} \bra{2 \beta}
\hat{B}^\dag_{n' k'} \ket{1 n'' k''}  \frac{1}{2}\ \frac{d}{dt'}
\Bigg( \langle \hat{B}_{n' k'} \left.\right>^{(1)}(t') \langle
\hat{B}_{n'' k''} \left.\right>^{(1)}(t') e^{-i \Omega(t-t')} \Bigg)
e^{+i \Omega(t-t')} = \nonumber\\
&& \hspace{-1.0cm} = \hbar \sum_{n' k';n'' k''} \int_{-\infty}^t dt'
\bra{1 \tilde n \tilde k} \hat{B}_{n k} \big(
\frac{\hat{H}_{c}}{\hbar} - \omega_{1 \tilde n \tilde k} - \omega_{1
n k} \big) e^{-i \frac{\hat{H}_{c}}{\hbar} (t-t')} \hat{B}^\dag_{n'
k'} \ket{1 n'' k''} e^{- \Gamma_{\text{xx}}(t-t')}
\nonumber \\
&& \hspace{0.0cm} \frac{1}{2}\ \frac{d}{dt'} \Bigg( \langle
\hat{B}_{n' k'} \left.\right>^{(1)}(t') \langle \hat{B}_{n'' k''}
\left.\right>^{(1)}(t') e^{-i \Omega(t-t')} \Bigg) e^{+i
\Omega(t-t')}\, ,\end{eqnarray} where in the last lines we have
resummed all the biexciton subspace by virtue of its completeness.
By performing an integration by part, Eq.\,(\ref{conti c 2}) can be
rewritten as
\begin{eqnarray}\label{conti c 3}
&& \hspace{-1.0cm} = \frac{1}{2}\hbar \!\!\! \sum_{n' k';n''
k''}\!\! \Bigg[ \Bigg\{ e^{i(\omega_{1 n' k'} + \omega_{1 n'' k''} -
2 i \Gamma_{\text{x}} + i \Gamma_{\text{xx}})(t-t')} \bra{1 \tilde n
\tilde k} \hat{B}_{n k} \big( \frac{\hat{H}_{c}}{\hbar} - \omega_{1
\tilde n \tilde k} - \omega_{1 n k} \big) e^{-i
\frac{\hat{H}_{c}}{\hbar} (t-t')}\!
\hat{B}^\dag_{n' k'}\!\! \ket{1 n'' k''} \nonumber \\
&& \langle \hat{B}_{n' k'} \left.\right>^{(1)}(t') \langle
\hat{B}_{n'' k''} \left.\right>^{(1)}(t') e^{-i \Omega(t-t')}
\Bigg\}^t_{-\infty}- \nonumber \\
&& - \int_{-\infty}^t dt' \langle \hat{B}_{n' k'}
\left.\right>^{(1)}(t') \langle \hat{B}_{n'' k''}
\left.\right>^{(1)}(t') e^{-i \Omega(t-t')} \frac{d}{dt'} \Bigg\{
e^{i(\omega_{1 n' k'} + \omega_{1 n'' k''} - 2 i \Gamma_{\text{x}} +
i \Gamma_{\text{xx}})(t-t')} \nonumber \\
&& \bra{1 \tilde n \tilde k} \hat{B}_{n k} \big(
\frac{\hat{H}_{c}}{\hbar} - \omega_{1 \tilde n \tilde k} - \omega_{1
n k} \big) e^{-i \frac{\hat{H}_{c}}{\hbar} (t-t')} \hat{B}^\dag_{n'
k'} \ket{1 n'' k''} \Bigg\} \Bigg]=\end{eqnarray}
\begin{eqnarray}\label{conti c 4}
&& \hspace{-1.0cm} = \frac{1}{2} \hbar \!\!\!\! \sum_{n' k';n''
k''}\!\! \Bigg\{ \bra{1 \tilde n \tilde k} \hat{B}_{n k} \big(
\frac{\hat{H}_{c}}{\hbar} - \omega_{1 \tilde n \tilde k} - \omega_{1
n k} \big) \! \hat{B}^\dag_{n' k'}\!\! \ket{1 n'' k''} \langle
\hat{B}_{n' k'} \left.\right>^{(1)}(t) \langle
\hat{B}_{n'' k''} \left.\right>^{(1)}(t) - \nonumber \\
&& - \int_{-\infty}^t dt' \langle \hat{B}_{n' k'}
\left.\right>^{(1)}(t') \langle \hat{B}_{n'' k''}
\left.\right>^{(1)}(t') e^{-i \Omega(t-t')} \frac{d}{dt'} \Bigg\{
e^{i(\omega_{1 n' k'} + \omega_{1 n'' k''} - 2 i \Gamma_{\text{x}} +
i \Gamma_{\text{xx}})(t-t')} \nonumber \\
&& \bra{1 \tilde n \tilde k} \hat{B}_{n k} \big(
\frac{\hat{H}_{c}}{\hbar} - \omega_{1 \tilde n \tilde k} - \omega_{1
n k} \big) e^{-i \frac{\hat{H}_{c}}{\hbar} (t-t')} \hat{B}^\dag_{n'
k'} \ket{1 n'' k''} \Bigg\}\, .\end{eqnarray}
The first and the second term can be expressed in terms of a double
commutator structure:
\begin{equation}\label{doppio comm 1}
\bra{1 \tilde n \tilde k} \hat{B}_{n k} \big(
\frac{\hat{H}_{c}}{\hbar} - \omega_{1 \tilde n \tilde k} - \omega_{1
n k} \big) = \bra{0} [\hat{B}_{\tilde n \tilde k},[\hat{B}_{n
k},\hat{H}_{c}]] \doteq \bra{0} \hat{D}_{\tilde n \tilde k,n k}\, ,
\end{equation} where a \textit{force} operator $\hat{D}$  is defined \cite{Sham PRL95}
and
\begin{eqnarray}\label{doppio comm 2}
&& \frac{d}{dt'} \Bigg\{ e^{i(\omega_{1 n' k'} + \omega_{1 n'' k''}
- 2 i \Gamma_{\text{x}} +
i \Gamma_{\text{xx}})(t-t')} \nonumber \\
&& \bra{1 \tilde n \tilde k} \hat{B}_{n k} \big(
\frac{\hat{H}_{c}}{\hbar} - \omega_{1 \tilde n \tilde k} - \omega_{1
n k} \big) e^{-i \frac{\hat{H}_{c}}{\hbar} (t-t')}
\hat{B}^\dag_{n' k'} \ket{1 n'' k''} \Bigg\} = \nonumber \\
&& = \frac{d}{dt'} \Bigg\{ \bra{0} \hat{D}_{\tilde n \tilde k,n k}
e^{-i \frac{\hat{H}_{c}}{\hbar} (t-t')} \hat{B}^\dag_{n' k'}
\hat{B}^\dag_{n'' k''} \ket{0} e^{i(\omega_{1 n' k'} + \omega_{1 n''
k''} - 2 i \Gamma_{\text{x}} + i \Gamma_{\text{xx}})(t-t')} \Bigg\}
= \nonumber \\
&& = \bra{0} \hat{D}_{\tilde n \tilde k,n k} e^{-i
\frac{\hat{H}_{c}}{\hbar} (t-t')} i \Big( \frac{\hat{H}_{c}}{\hbar}
- \omega_{1 n' k'} - \omega_{1 n'' k''}
- i (\Gamma_{\text{xx}} - 2 \Gamma_{\text{x}}) \Big) \nonumber \\
&& \hspace{1.0cm} \hat{B}^\dag_{n' k'} \hat{B}^\dag_{n'' k''}
\ket{0} e^{i(\omega_{1 n' k'} + \omega_{1 n'' k''} - 2 i
\Gamma_{\text{x}} + i \Gamma_{\text{xx}})(t-t')} = \nonumber\\
&& = e^{i(\omega_{1 n' k'} + \omega_{1 n'' k''} - 2 i
\Gamma_{\text{x}} + i \Gamma_{\text{xx}})(t-t')} i
F^{n'' k'', n' k'}_{\tilde n \tilde k,n k}(t-t') + \\
&& + (\Gamma_{\text{xx}} - 2 \Gamma_{\text{x}})e^{i(\omega_{1 n' k'}
+ \omega_{1 n'' k''} - 2 i \Gamma_{\text{x}} + i
\Gamma_{\text{xx}})(t-t')} \bra{0} \hat{D}_{\tilde n \tilde k,n
k}(t-t')\hat{B}^\dag_{n' k'} \hat{B}^\dag_{n'' k''} \ket{0}
\nonumber\, , \end{eqnarray} where the memory kernel reads
\begin{equation}\label{F}
F^{n'' k'', n' k'}_{\tilde n \tilde k,n k}(t-t') = \bra{0}
\hat{D}_{\tilde n \tilde k,n k}(t-t') \hat{D}^\dag_{n'' k'',n' k'}
\ket{0}\, . \end{equation}
The usual time dependence in the Heisenberg picture is given by $
\hat{D}(\tau) = e^{i (\hat{H}_{c}/\hbar) \tau}\hat{D}e^{-i
(\hat{H}_{c}/\hbar) \tau}$. Altogether, the nonlinear term
originating from Coulomb interaction can be written as
\begin{eqnarray}\label{H-F}
&& \hspace{-1.0cm} Q^{\text{COUL}(2)}_{n k;\tilde n \tilde k} =
\sum_{\beta}c^{(1)}_{n k;\tilde n \tilde k;\beta}
\langle \hat{X}_{0; 2 \beta} \rangle^{(2)} = \nonumber \\
&& \hspace{-1.0cm} \frac{1}{2} \hbar \sum_{n' k';n'' k''} \Bigg\{
\bra{0} \hat{D}_{\tilde n \tilde k,n k} \hat{B}^\dag_{n' k'}
\hat{B}^\dag_{n'' k''} \ket{0} \langle \hat{B}_{n' k'}
\left.\right>^{(1)}(t) \langle \hat{B}_{n''
k''} \left.\right>^{(1)}(t) - \nonumber \\
&& - i \int_{-\infty}^t dt' F^{n'' k'', n' k'}_{\tilde n \tilde k,n
k}(t-t') \langle \hat{B}_{n' k'} \left.\right>^{(1)}(t') \langle
\hat{B}_{n'' k''} \left.\right>^{(1)}(t') e^{-
\Gamma_{\text{xx}}(t-t')} \Bigg\} - \\
&& \hspace{-1.0cm} - \frac{\hbar}{2}(\Gamma_{\text{xx}} - 2
\Gamma_{\text{x}}) \hspace{-0.1cm} \sum_{\substack{ n' k'\\n'' k''}}
\hspace{-0.1cm} \int_{-\infty}^t dt' \bra{0} \hat{D}_{\tilde n
\tilde k,n k}(t-t')\hat{B}^\dag_{n' k'} \hat{B}^\dag_{n'' k''}
\ket{0} \langle \hat{B}_{n' k'} \left.\right>^{(1)}(t') \langle
\hat{B}_{n'' k''} \left.\right>^{(1)}(t') \nonumber \, .
\end{eqnarray}
For later purpose we are interested in the optical response dominate
by the 1S exciton sector, with $\Gamma_{\text{xx}} \simeq 2
\Gamma_{\text{x}}$ in the cases of counter- and co-circularly
polarized waves. Specifying to this case the Coulomb-induced term
with Eq.\, (\ref{H-F}) becomes
\begin{eqnarray}\label{Coul 1S}
&& \frac{d}{dt}\left<\right. \hat{B}_{\pm {\bf k}}
\left.\right>^{+(3)}\Biggl|_{\text{COUL}}= -i \bar{\omega}_{\bf k}
\left<\right. \hat{B}_{\pm {\bf k}}
\left.\right>^{+(3)}-\frac{i}{\hbar} \sum_{\tilde{\sigma} {\bf
\tilde{k}}} \langle \hat{B}_{\tilde \sigma {\bf \tilde k}}
\rangle^{*(1)} Q^{\text{COUL}(2)}_{\pm {\bf k};\tilde \sigma {\bf
\tilde k}} = \\
&& =-i \bar{\omega}_{\bf k} \left<\right. \hat{B}_{\pm {\bf k}}
\left.\right>^{+(3)}-\frac{i}{\hbar} \sum_{\bf k' k'' \tilde{k}}
\delta_{\bf k+\tilde k;k'+k''} V_{\text{xx}} \langle \hat{B}_{\pm
{\bf \tilde k}} \rangle^{*(1)}(t) \langle \hat{B}_{\pm {\bf k'}}
\rangle^{(1)}(t) \langle \hat{B}_{\pm {\bf
k''}}\rangle^{(1)}(t) + \nonumber \\
&& -\frac{1}{\hbar} \sum_{\substack{\sigma' \sigma'' \tilde{\sigma}\\
\bf k' k'' \tilde{k}}} \delta_{\bf k+\tilde k;k'+k''}
\delta_{\pm+\tilde \sigma;\sigma'+\sigma''} \langle \hat{B}_{\tilde
\sigma {\bf \tilde k}} \rangle^{*(1)}(t)  \nonumber \\
&& \hspace{3.0cm} \int_{-\infty}^t dt' F^{\sigma' \sigma''}(t-t')
\langle \hat{B}_{\sigma' {\bf k}'} \left.\right>^{(1)}(t') \langle
\hat{B}_{\sigma'' {\bf k}''} \left.\right>^{(1)}(t') e^{-
\Gamma_{\text{xx}}(t-t')}\, , \nonumber
\end{eqnarray}
where, in order to lighten the notation, we dropped the two spin
indexes $\sigma$ and $\tilde \sigma$ in the four-particle kernel
function $F$ defined in Eq.\, (\ref{F}) for they are already
univocally determined once chosen the others (i.e. $\sigma'$ and
$\sigma''$) as soon as their selection rule ($\delta_{\sigma+\tilde
\sigma;\sigma'+\sigma''}$) is applied. Moreover, the $\hbar/2$ has
been reabsorbed in the Coulomb nonlinear coefficients
$V_{\text{xx}}$ and $F^{\sigma' \sigma''}(t-t')$. A detail
microscopic account for the mean-field $V_{\text{xx}}$, for the
$F$'s and their selection rules are considered in \cite{Takayama
EPJ, Kwong-Binder PRB 2001}. For the range of ${\bf k}$-space of
interest, i.e. $|{\bf k}| \ll \frac{\pi}{a_{\text{x}}}$ (much lower
than the inverse of the exciton Bohr radius) they are largely
independent on the center of mass wave vectors. While
$V_{\text{xx}}$ and $F^{\pm \pm}(t-t')$ (i.e. co-circularly
polarized waves) conserve the polarizations, $F^{\pm \mp}(t-t')$ and
$F^{\mp \pm }(t-t')$ (counter-circular polarization) give rise to a
mixing between the two circularly polarizations. The physical origin
of the three terms in Eq.\, (\ref{H-F}) can be easily understood:
the first is the Hartee-Fock or mean-field term representing the
first order treatment in the Coulomb interaction between excitons,
the second term is a pure biexciton (four-particle correlations)
contribution. This coherent memory may be thought as a non-Markovian
process involving the two-particle (excitons) states interacting
with a bath of four-particle correlations \cite{Sham PRL95}.
%The analytical expression of the kernel function $F(t)$ in terms of the
%two-exciton wave functions can be found in Refs.\, \cite{Takayama
%EPJ, Kwong-Binder PRB 2001}.
Equation\, (\ref{H-F}) even if formally similar to that of Ref.
\cite{Sham PRL95}, represents its extension including polaritonic
effects due to the presence of the cavity . It has been possible
thanks to the inclusion of the dynamics of the cavity modes whereas
in Ref.\, \cite{Sham PRL95} the electromagnetic field entered as a
parameter only. Former analogous extensions have been obtained
within a semiclassical model \cite{Takayama EPJ, Kwong-Binder PRB
2001,Savasta PRL2003}. The strong exciton-photon coupling does not
modify the memory kernel because four-particle correlations do not
couple directly to cavity photons. As pointed out clearly in Ref.\,
\cite{Savasta PRL2003}, cavity effects alter the phase dynamics of
two-particle states during collisions, indeed, the phase of
two-particle states in SMCs oscillates with a frequency which is
modified respect to that of excitons in bare QWs, thus producing a
modification of the integral in Eq.\, (\ref{H-F}). In this way the
exciton-photon coupling $V_{n k}$ affects the exciton-exciton
collisions that govern the polariton amplification process. Ref.
\cite{Savasta PRL2003} considers the first (mean-field) and the
second (four-particle correlation) terms in the particular case of
cocircularly polarized waves, calling them without indexes as
$V_{\text{xx}}$ and $F(t)$ respectively. In Fig.\, 1 they show
${\cal F}(\omega)$, the Fourier transform of $F(t)$ plus the
mean-field term $V_{\text{xx}}$,
\begin{equation}
{\cal F}(\omega) = V_{\text{xx}} -i \int^\infty_{- \infty} dt F(t)
e^{i \omega t}\, .\end{equation} Its imaginary part is responsible
for the frequency dependent excitation induced dephasing, it
reflects the density of the states of two-exciton pair coherences.
Towards the negative detuning region the dispersive part Re$({\cal
F})$ increases whereas the absorptive part Im$({\cal F})$ goes to
zero. The former comprises the mean-field contribution effectively
reduced by the four-particle contribution. Indeed, the figure shows
the case with a binding energy of 13.5 meV, it gives $V_{\text{xx}}
n_{\text{sat}} \simeq 11.39$ meV which clearly is an upper bound for
Re$({\cal F})$ for negative detuning. The contribution carried by
$F(t)$ determines an effective reduction of the mean-field
interaction (through its imaginary part which adds up to
$V_{\text{xx}}$) and an excitation induced dephasing. It has been
shown \cite{Savasta PRL2003} that both effects depends on the sum of
the energies of the scattered polariton pairs. The third term in
Eq.\, (\ref{H-F}) can be thought as a reminder of the mismatch
between the picture of a biexciton as a composite pair of exciton.
In the following we will set $\Gamma_{\text{xx}} \simeq 2
\Gamma_{\text{x}}$.

The other nonlinear source term in Eq.\, (\ref{R nl}) depends
directly on the exciton wave function and reads
\begin{equation}
\sum_{\tilde{n} \tilde{k}} \langle \hat{B}_{\tilde n \tilde k}
\rangle^{*(1)} \sum_{n' k',n'' k''}C^{n' k',n'' k''}_{\tilde n
\tilde k,n k} \langle \hat{a}_{k'} \rangle^{(1)} \langle
\hat{B}_{n'' k''} \rangle^{(1)}\, .\end{equation} It represents a
phase-space filling (PSF) contribution, due to the Pauli blocking of
electrons. It can be developed as follows,
\begin{eqnarray}\label{PSFconti}
&& C^{n' k',n'' k''}_{\tilde n \tilde k,n k} = V_{n' k'} \bra{1
\tilde{n} \tilde \sigma {\bf \tilde{k}}} \delta_{(n' k');(n k)} -
[\hat{B}_{n \sigma {\bf k}}, \hat{B}^\dag_{n' \sigma' {\bf k'}} ] \ket{1 n'' \sigma'' {\bf k''}} = \nonumber \\
&& = V_{n' k'} \delta_{\sigma,\sigma'} \Biggl\{ \sum_{\bf q}
\Phi^{{\bf k}\, *}_{n \sigma {\bf q}} \Phi^{\bf k'}_{n' \sigma'
({\bf q}+\eta_h({\bf k'}-{\bf k}))} \bra{1 \tilde{n} \tilde \sigma
{\bf \tilde{k}}} \hat{c}^\dag_{\sigma',{\bf q}+\eta_h({\bf k'}-{\bf
k})+\eta_e{\bf k'}} c_{\sigma,{\bf q}+\eta_e{\bf k}} \ket{1 n'' \sigma'' {\bf k''}} + \nonumber \\
&& \sum_{\bf q} \Phi^{{\bf k}\, *}_{n \sigma {\bf q}} \Phi^{\bf
k'}_{n' \sigma' ({\bf q}-\eta_e({\bf k'}-{\bf k}))} \bra{1 \tilde{n}
\tilde \sigma {\bf \tilde{k}}} \hat{d}^\dag_{\sigma',-{\bf
q}+\eta_e({\bf k'}-{\bf k})+\eta_h{\bf k'}} d_{\sigma,-{\bf
q}+\eta_h{\bf k}} \ket{1 n'' \sigma'' {\bf k''}} \Biggr\} = \nonumber \\
&& = V_{n' k'} \delta_{\sigma,\sigma'} \delta_{\bf k+\tilde
k;k'+k''} \Biggl\{ \sum_{\bf q} \Phi^{{\bf k}\,
*}_{n \sigma {\bf q}} \Phi^{\bf k'}_{n' \sigma' {\bf q}_0} \Phi^{{\bf \tilde k}\,
*}_{{\tilde{n} \tilde \sigma} {\bf q}_1} \Phi^{\bf k''}_{n'' \sigma'' {\bf q}_2} + \nonumber \\
&& \sum_{\bf q} \Phi^{{\bf k}\, *}_{n \sigma {\bf q}} \Phi^{\bf
k'}_{n' \sigma' {\bf q}_3} \Phi^{{\bf \tilde k}\,
*}_{{\tilde{n} \tilde \sigma} {\bf q}_4} \Phi^{\bf k''}_{n'' \sigma'' {\bf q}_5}
\Biggr\} \, , \end{eqnarray}
the explicit expressions of the ${\bf q}$'s are given in \cite{q's}.

Thus, the nonlinear dynamics of Eq.\, (\ref{dt B up to 3fin}) driven
by $\hat{H}_I$ can be written
\begin{eqnarray}\label{dt B H_I}
&&\frac{d}{dt}\left<\right. \hat{B}_{n \sigma {\bf k}}
\left.\right>^{+(3)}\Bigl|_{\hat{H}_I} = +i \frac{V_{n \sigma {\bf
k}}}{\hbar} \left<\right. \hat{a}_{\sigma {\bf k}}
\left.\right>^{+(3)} -\frac{i}{\hbar} \sum_{\substack{n' n''
\tilde{n}\\ \bf k' k'' \tilde{k}}} \delta_{\bf k+\tilde k;k'+k''}
\langle \hat{B}_{\tilde n \sigma {\bf \tilde k}}
\rangle^{*(1)}  \nonumber \\
&& \hspace{1.0cm} \langle \hat{a}_{\sigma {\bf k'}} \rangle^{(1)}
\langle \hat{B}_{n'' \sigma {\bf k''}}\rangle^{(1)} \tilde
V^*_{\sigma}  \Bigl[ \sum_{\bf q} \Phi^{{\bf k}\, *}_{n \sigma {\bf
q}} \Phi^{{\bf \tilde k}\, *}_{{\tilde{n} \sigma} {\bf q}_1}
\Phi^{\bf k''}_{n'' \sigma {\bf q}_2} + \sum_{\bf q} \Phi^{{\bf k}\,
*}_{n \sigma {\bf q}} \Phi^{{\bf \tilde k}\, *}_{{\tilde{n} \sigma} {\bf q}_4} \Phi^{\bf k''}_{n'' \sigma {\bf q}_5}
\Bigr] \, .\end{eqnarray}
We are interested in studying polaritonic effects in SMCs where the
optical response involves mainly excitons belonging to the 1S band
with wave vectors close to normal incidence, i.e. $|{\bf k}| \ll
\frac{\pi}{a_{\text{x}}}$ (much lower than the inverse of the
exciton Bohr radius). In this case the exciton relative wave
functions are independent on spins as well as on the center of mass
wave vector. They are such that $\sum_{\bf q=-\infty}^\infty
|\Phi_{\bf q}|^2 = 1$, i.e. $\Phi_{\bf q} = \frac{1}{\sqrt{A}}
\frac{\sqrt{2 \pi} 2 a_{\text{x}}}{(1+(a_{\text{x}}|{\bf
q}|)^2)^{3/2}}$, $a_{\text{x}}$ is the exciton Bohr radius. From now
on whenever no excitonic level is specified the 1S label is understood.
It yields
\begin{eqnarray}\label{dt B H_I 2}
&&\frac{d}{dt}\left<\right. \hat{B}_{\sigma {\bf k}}
\left.\right>^{+(3)}\Bigl|_{\hat{H}_I} = +i \frac{V_{\sigma {\bf
k}}}{\hbar} \left<\right. \hat{a}_{\sigma {\bf k}}
\left.\right>^{+(3)} -\frac{i}{\hbar} \sum_{\bf k' k'' \tilde{k}} \delta_{\bf k+\tilde k;k'+k''} \nonumber \\
&& \hspace{1.0cm} \langle \hat{B}_{\tilde \sigma {\bf \tilde k}}
\rangle^{*(1)} \langle \hat{a}_{\sigma {\bf k'}} \rangle^{(1)}
\langle \hat{B}_{\sigma {\bf k''}}\rangle^{(1)} 2 \tilde
V^*_{\sigma} O^{\text{PSF}}\, ,\end{eqnarray}
where the overlap $O^{\text{PSF}}$ has been calculated in the case
of zero center of mass wave vector, namely
\begin{equation}\label{O^PSF}
O^{\text{PSF}} = \sum_{\bf q} \Phi^{*}_{{\bf q}} \Phi^{*}_{\bf q}
\Phi_{\bf q}\nonumber \, .\end{equation}
In SMCs a measured parameter is the so-called vacuum Rabi splitting
$V_{n \sigma {\bf k}}$ \cite{Baumberg} of the 1S excitonic
resonance, for the range of ${\bf k}$-space of interest essentially
constant. Defining $V \doteq V_{\sigma} = \tilde V_{\sigma} \sqrt{A}
\phi^*(0)$
\begin{equation}
\tilde V^*_{\sigma} O^{\text{PSF}} = \frac{V}{\sqrt{A}\phi^*(0)}
O^{\text{PSF}} = \frac{8}{7} \frac{\pi a^2_{\text{x}}}{A}V =
\frac{1}{2} \frac{V}{n_\text{sat}}\, ,\end{equation} where we have
set $n_{\text{sat}} \doteq (7/16)\!\! \cdot \!\! (A/\pi
a^2_{\text{x}})$, called saturation density.

In terms of the two circular polarizations the dynamics induced by
$\hat{H}_I$ finally reads
\begin{equation}\label{dt B H_I fin}
\frac{d}{dt}\left<\right. \hat{B}_{\pm {\bf k}}
\left.\right>^{+(3)}\Bigl|_{\hat{H}_I} = +i \frac{V}{\hbar}
\left<\right. \hat{a}_{\pm {\bf k}} \left.\right>^{+(3)}
-\frac{i}{\hbar} \sum_{\bf \tilde{k}} \langle \hat{B}_{\pm {\bf
\tilde k}} \rangle^{*(1)} Q^{\text{PSF}(2)}_{\pm {\bf k};\tilde
\sigma {\bf \tilde k}}\, ,\end{equation}
where
\begin{equation}\label{PSF term}
\sum_{\bf \tilde{k}} \langle \hat{B}_{\pm {\bf \tilde k}}
\rangle^{*(1)} Q^{\text{PSF}(2)}_{\pm {\bf k};\tilde \sigma {\bf
\tilde k}} = \frac{V}{n_\text{sat}} \sum_{\bf k' k'' \tilde{k}}
\delta_{\bf k+\tilde k;k'+k''} \langle \hat{B}_{\pm {\bf \tilde k}}
\rangle^{*(1)} \langle \hat{a}_{\pm {\bf k'}} \rangle^{(1)} \langle
\hat{B}_{\pm {\bf k''}}\rangle^{(1)}\, .\end{equation}
The same lines of argument can be followed for computing the
Coulomb-induced interactions $Q^{\text{COUL}(2)}$ \cite{Takayama
EPJ, Kwong-Binder PRB 2001}.

We are lead to introduce the saturation density for two main
reasons. The most obvious is our interest to refer this work to the
literature where $n_{\text{sat}}$ is extensively used \cite{Langbein
PRB2004,Savasta PRL2003,Ciuti SST,Savasta PRB2001}. The other most
interesting reason is that we can directly compute this quantity.
Indeed, the equation of motion for the exciton operator reads
\begin{eqnarray}\label{dt B PSF }
&& \frac{d}{dt}\left<\right. \hat{B}_{\pm {\bf k}}
\left.\right>^{+(3)}= -i \bar{\omega}_{\bf k} \left<\right.
\hat{B}_{\pm {\bf k}} \left.\right>^{+(3)} +i \frac{V}{\hbar}
\left<\right. \hat{a}_{\pm {\bf k}} \left.\right>^{+(3)}
-\frac{i}{\hbar} \sum_{\tilde{\sigma}=\pm {\bf \tilde{k}}} \langle
\hat{B}_{\tilde \sigma {\bf \tilde k}} \rangle^{*(1)}
Q^{\text{COUL}(2)}_{\pm {\bf k};\tilde \sigma {\bf
\tilde k}} \nonumber \\
&& \hspace{1.0cm} -\frac{i}{\hbar} 2 \frac{V}{\sqrt{A}\phi^*(0)}
O^{\text{PSF}} \sum_{\bf k' k'' \tilde{k}} \delta_{\bf k+\tilde
k;k'+k''} \langle \hat{B}_{\pm {\bf \tilde k}} \rangle^{*(1)}
\langle \hat{a}_{\pm {\bf k'}} \rangle^{(1)} \langle \hat{B}_{\pm
{\bf k''}}\rangle^{(1)} \nonumber\, . \end{eqnarray}
Leaving apart the discrepancy between the order in the DCTS we can
compute the so-called \textit{oscillator strength} (OS), defined as
what multiplies the photon expectation values $\langle \hat{a}_{\pm
{\bf k}=0} \rangle$,
\begin{eqnarray}\label{OS}
OS =\!\! i\frac{V}{\hbar} {\Bigg (} 1 - \frac{2}{\sqrt{A}\phi^*(0)}
O^{\text{PSF}}\Big[ \langle \hat{B}_{\pm 0} \rangle^{*(1)} \langle
\hat{B}_{\pm 0} \rangle^{(1)} \Big] {\Bigg )}\, .
\end{eqnarray} The saturation density may be defined as the exciton density that
makes the oscillator strength to be zero. We obtain
\begin{equation}\label{nsat value}
n_{\text{sat}} = \Biggl( \frac{2}{\sqrt{A}\phi^*(0)} O^{\text{PSF}}
\Biggr)^{-1} = \frac{A}{\pi a^2_{\text{x}}} \ \frac{7}{16}\, .
\end{equation}

Eventually, the lowest order ($\chi^{(3)}$) nonlinear optical
response in SMCs are described by the following set of coupled
equations:
\begin{equation}\label{dt a}
\frac{d}{dt}\left<\right. \hat{a}_{\pm {\bf k}} \left.\right>^{+(3)}
= -i \bar{\omega}^c_{\bf k} \left<\right. \hat{a}_{\pm {\bf k}}
\left.\right>^{+(3)} + i \frac{V}{\hbar} \left<\right. \hat{B}_{\pm
{\bf k}} \left.\right>^{+(3)} + t_c \frac{E_{\pm {\bf k}}}{\hbar}\,
, \end{equation}
\begin{equation}\label{dt B}
\frac{d}{dt}\left<\right. \hat{B}_{\pm {\bf k}} \left.\right>^{+(3)}
= -i \bar{\omega}_{\bf k} \left<\right. \hat{B}_{\pm {\bf k}}
\left.\right>^{+(3)} +i \frac{V}{\hbar} \left<\right. \hat{a}_{\pm
{\bf k}} \left.\right>^{+(3)} -\frac{i}{\hbar} \sum_{\tilde{\sigma}
{\bf \tilde{k}}} \langle \hat{B}_{\tilde \sigma {\bf \tilde k}}
\rangle^{*(1)} R^{(2)}_{\pm {\bf k};\tilde \sigma {\bf \tilde k}}\,
,\end{equation}
with $\sum_{\tilde{\sigma} {\bf \tilde{k}}} \langle \hat{B}_{\tilde
\sigma {\bf \tilde k}} \rangle^{*(1)} R^{(2)}_{\pm {\bf k};\tilde
\sigma {\bf \tilde k}} = \sum_{\tilde{\sigma} {\bf \tilde{k}}}
\langle \hat{B}_{\tilde \sigma {\bf \tilde k}} \rangle^{*(1)}
Q^{\text{COUL}(2)}_{\pm {\bf k};\tilde \sigma {\bf \tilde k}} +
\sum_{\bf \tilde{k}} \langle \hat{B}_{\pm {\bf \tilde k}}
\rangle^{*(1)} Q^{\text{PSF}(2)}_{\pm {\bf k};\tilde \sigma {\bf
\tilde k}}$, with the first of the two addenda originating from
Coulomb interaction, Eq.\, (\ref{Coul 1S}), whereas the second
represents the phase-space filling contribution written in Eqs.
(\ref{PSF term}). Starting from here, in the strong coupling case,
it might be useful to transform the description into a polariton
basis. The proper inclusion of dephasing/relaxation and the
application of these equations to parametric processes, in the
strong coupling regime, is described in another paper of ours
\cite{nostro PRB}.

Equations\, (\ref{dt a}) and (\ref{dt B}) is exact to the third
order in the exciting field. While a systematic treatment of
higher-order optical nonlinearities would require an extension of
the equations of motions (see e.g. Appendix), a restricted class of
higher-order effects can be obtained from solving equations\,
(\ref{dt a}) and (\ref{dt B}) self-consistently up to arbitrary
order as it is usually employed in standard nonlinear optics. This
can be simply accomplished by replacing, in the nonlinear sources,
the linear excitonic polarization and light fields with the total
fields \cite{Sham PRL95,Savasta PRL2003,Buck}. Multiple-scattering
processes are expected to be very effective in cavity-embedded QW's
due to multiple reflections at the Bragg mirrors.

\section{Parametric Photoluminescence: Towards Semiconductor Quantum
Optics}\label{4}

Entanglement is one of the key features of quantum information and
communication technology \cite{Nielsen-Chuang}. Parametric
down-conversion is the most frequently used method to generate
highly entangled pairs of photons for quantum-optics applications,
such as quantum cryptography and quantum teleportation. This
$\chi^{(3)}$ optical nonlinear process consists of the scattering of
two polaritons generated by a coherent pump beam into two final
polariton modes. The total energy and momentum of the final pairs
equal that of pump polariton pairs. The scattering can be
spontaneous (parametric emission) or stimulated by a probe beam
resonantly exciting one of the two final polariton modes. In 2005 an
experiment probing quantum correlations of (parametrically emitted)
cavity polaritons by exploiting quantum complementarity has been
proposed and realized \cite{Savasta PRL2005}. The most common set-up
for parametric emission is the one where a single  coherent pump
feed resonantly excites the structure at a given energy and wave
vector, $\bf{k}_p$. Within the DCTS we shall employ Eqs.\, (\ref{dt
a up to 3}), (\ref{dt B up to 3}) and Eq.\, (\ref{X02}) in
operatorial form, provided all the equations to become fully
significant as soon as the expectation value quantities we shall
work out would lie within the consistent perturbative DCTS order we
set from the beginning \cite{HRS Savasta}. In order to be more
\textit{specific} we shall derive explicitly the case of input light
beams activating only the $1 S$ exciton sector with all the same
circularly (e.g. $\sigma^+$) polarization, thus excluding the
coherent excitation of bound two-pair coherences (biexciton) mainly
responsible for polarization-mixing \cite{Sham PRL95}. Equations
involving polariton pairs with opposite polarization can be derived
in complete analogy following the same steps. Starting from the
Heisenberg equations for the exciton and photon operators and
keeping only terms providing lowest order nonlinear response (in the
input light field) we obtain,
\begin{equation}\label{dt Y up to 3}
\frac{d}{dt} \hat{a}_{k}  = -i \omega^c_k  \hat{a}_{k} + i
\frac{V^*_{k}}{\hbar} \hat{B}_{k} + t_c \frac{E_k}{\hbar}\, ,
\end{equation}
\begin{eqnarray}\label{dt X up to 3}
&&\frac{d}{dt} \hat{B}_{k} = - i \omega_{k} \hat{B}_{k}
+i \frac{V_{k}}{\hbar}\ \hat{a}_{k} + \nonumber \\
&& \hspace{2.0cm} + \frac{i}{\hbar} \sum_{\tilde{k}, k', \alpha}
V_{k'} \bra{1 \tilde{k}} [\hat{B}_{k}, \hat{B}^\dag_{k'} ] -
\delta_{(k'),(k)} \ket{1 \alpha} \hat{X}_{1 \tilde{k},0}
\hat{X}_{0,1 \alpha} \hat{Y}_{0; 1
k'} - \nonumber \\
&& \hspace{3.0cm} - \frac{i}{\hbar} \sum_{\tilde{k} \beta}(\omega_{2
\beta} - \omega_{1 \tilde k} - \omega_{1 k}) \bra{1 \tilde k}
\hat{B}_{k} \ket{2 \beta} \hat{X}_{1 \tilde{k},0} \hat{X}_{0, 2
\beta} \, .\end{eqnarray} In the following we will assume that the
pump polaritons driven by a quite strong coherent input field
consists of a classical ($\mathbb{C}$-number) field. This
approximation is in close resemblance to the two approximations
performed in the previous section (linearization of fluctuations and
coherent nonlinear processes). We shall show that under this
approximation, we may perform the same manipulations ending up to a
set of coupled equations analogous to Eqs.\, (\ref{dt a}) and
(\ref{dt B}). In addition, having a precise set-up chosen, we will
be able to specialize our equations and give an explicit account of
the parametric contributions as well as the shifts the lowest order
nonlinear dynamics provides. We shall retain only those terms
containing the semiclassical pump amplitude at $k_p$ twice, thus
focusing on the ``direct" pump-induced nonlinear parametric
scattering processes. It reads

\begin{eqnarray}\label{dt X con shift}
&&\frac{d}{dt} \hat{B}_{\pm {\bf k}} = - \omega_{\bf k}
\hat{B}_{\pm {\bf k}} + i \frac{V}{\hbar}\ \hat{a}_{\pm {\bf k}} - \\
&& \hspace{0.0cm} - \frac{i}{\hbar} \frac{V}{n_{\text{sat}}}
\sum_{\bf \tilde{k}, k', k''} \delta_{\bf k+\tilde{k},k'+k''}
\hat{X}_{1 \pm {\bf \tilde{k}},0} \hat{X}_{0,1 \pm{\bf k''}}
\hat{Y}_{0; 1 \pm {\bf k'}} ( \delta_{{\bf k''},{\bf k}_p}
\delta_{{\bf k'},{\bf k}_p} + \delta_{{\bf \tilde{k}},{\bf k}_p}
\delta_{{\bf k''},{\bf k}_p} + \delta_{{\bf \tilde{k}},{\bf k}_p} \delta_{{\bf k'},{\bf k}_p} ) - \nonumber \\
&& \hspace{0.0cm} - \frac{i}{\hbar} \sum_{\tilde \sigma {\bf
\tilde{k}}, \sigma_{\beta} {\bf k}_\beta} (\omega_{2 {\bf k}_\beta}
- \omega_{1 {\bf \tilde k}} - \omega_{1 {\bf k}}) \bra{1 {\bf \tilde
\sigma \tilde k}} \hat{B}_{\pm {\bf k}} \ket{2 \sigma_{\beta} {\bf
k}_\beta} \hat{X}_{1 \tilde \sigma {\bf \tilde{k}},0} \hat{X}_{0, 2
\sigma_{\beta} {\bf k}_\beta} (\delta_{{k}_\beta,2 {k}_p} +
\delta_{{\tilde{k}},{k}_p}\delta_{{k}_\beta,{k}+{k}_p} )\nonumber\,
,
\end{eqnarray}
where we have already manipulated the phase-space filling matrix
element. Here in brackets the first addendum of each line would be
responsible for the parametric contribution, whereas the others will
give the shifts. It is understood, from now on, that the pump-driven
terms (e.g. the $X$ and $Y$ at $k_p$) are ${\mathbb{C}}$-numbers
coherent amplitudes like the semiclassical electromagnetic pump
field, we will make such distinction in marking with a ``hat" the
operators only.

We need some care in manipulating the Coulomb-induced terms, the
last line. Written explicitly it is

\begin{eqnarray}\label{X01 Coulomb}
&&\frac{d}{dt} \hat{B}_{\pm {\bf k}}{\Bigg |}_{\text{Coul}} = \nonumber \\
&& \hspace{0.5cm} - \frac{i}{\hbar} \sum_{\tilde \sigma {\bf
\tilde{k}}, \sigma_{\beta} {\bf k}_\beta} (\omega_{2 {\bf k}_\beta}
- \omega_{1 {\bf \tilde k}} - \omega_{1 {\bf k}}) \bra{1 \tilde
\sigma {\bf \tilde k}} \hat{B}_{\pm {\bf k}} \ket{2 \sigma_{\beta}
{\bf k}_\beta}
\hat{X}_{1 \tilde \sigma {\bf \tilde{k}},0} {X}_{0, 2 \sigma_p {\bf k}_p} + \nonumber \\
&& \hspace{0.5cm} - \frac{i}{\hbar} \sum_{\tilde \sigma {\bf
\tilde{k}}, \sigma_{\beta} {\bf k}_\beta} (\omega_{2 {\bf k}_\beta}
- \omega_{1 {\bf \tilde k}} - \omega_{1 {\bf k}}) \bra{1 \tilde
\sigma  {\bf \tilde k}} \hat{B}_{\pm {\bf k}} \ket{2 \sigma_{\beta}
{\bf k}_\beta} X_{1 \sigma_p {\bf k}_p,0} \hat{X}_{0, 2 \sigma_{{\bf
k}+{\bf k}_p}({\bf k}+{\bf k}_p)} \,\end{eqnarray}

As for the term containing ${X}_{0, 2 k_p}$, we are facing a
${\mathbb{C}}$-number which gives no problem in performing the very
same procedure of the previous chapter. As for the other we would
exploit the formal biexciton solution
\begin{eqnarray}\label{X02 sol}
&& \hat{X}_{0;2 (k+k_p)} (t) = \int_{-\infty}^t dt' e^{-i \omega_{2
(k+k_p)}(t-t')} \frac{i}{\hbar} {\Bigg (} V_{k_p} \bra{2 (k+k_p)}
\hat{B}^\dag_{k_p} \ket{1 k} \hat{X}_{0,1 k} Y_{0,1 k_p} + \nonumber
\\
&& \hspace{0.5cm} V_{k} \bra{2 (k+k_p)} \hat{B}^\dag_{k} \ket{1 k_p}
X_{0,1 k_p} \hat{Y}_{0,1 k} {\Bigg )} \, ,\end{eqnarray} where, for
the sake of consistence, we are neglecting $\hat{X}_{0;2 (k+k_p)}
(-\infty)$ because the biexciton, within the present approximations,
is always generated by an operator at $k$ times a classical
amplitude at $k_p$ which is always zero before the electromagnetic
impulse arrived. Moreover, an analogous identity such that of Eq.\,
(\ref{Sham identity}) is valid in the present context, namely
\begin{eqnarray}\label{Sham id per ops}
&& \frac{d}{dt} \Bigg( \hat{X}_{0,1 k} {X}_{0, 1 k_p} e^{-i
(\omega_{1 k} + \omega_{1 k_p})(t-t')} \bigg) =  \\
&& \hspace{2.5cm} = \bigg( i \frac{V_k}{\hbar} \hat{Y}_{0,1 k}
X_{0,1 k_p} + i \frac{V_{k_p}}{\hbar} Y_{0,1 k_p} \hat{X}_{0,1 k}
\bigg)e^{-i (\omega_{1 k} + \omega_{1 k_p})(t-t')} \nonumber\, .
\end{eqnarray}
With these tools at hand we are able to perform step by step the
manipulations of the previous section for all the quantities at
play. The final result reads
\begin{eqnarray}\label{dt X completa1}
&&\frac{d}{dt} \hat{B}_{\pm {\bf k}} = - \omega_{\bf k}
\hat{B}_{\pm {\bf k}} + i \frac{V}{\hbar}\ \hat{a}_{\pm {\bf k}} - \nonumber \\
&& \hspace{0.5cm} - \frac{i}{\hbar} \frac{V}{n_{\text{sat}}} \bigg(
\hat{X}_{1 \pm {\bf k}_i,0} X_{0,1 \pm {\bf k}_p} Y_{0, 1 \pm {\bf
k}_p} + X_{1 \pm {\bf k}_p,0} X_{0,1 \pm {\bf k}_p} \hat{Y}_{0, 1
\pm {\bf k}} + X_{1 \pm {\bf k}_p,0} \hat{X}_{0,1 \pm
{\bf k}} Y_{0, 1 \pm {\bf k}_p} \bigg) - \nonumber \\
&& \hspace{0.5cm} - \frac{i}{\hbar} \hat{X}_{1 \pm {\bf k}_i,0}(t)
\Bigg\{ V_{\text{xx}} X_{0,1 \pm {\bf k}_p}(t) X_{0,1 \pm {\bf
k}_p}(t) - i \int_{-\infty}^t dt' F^{\pm \pm}(t-t') X_{0,1 \pm {\bf
k}_p}(t')X_{0,1 \pm {\bf k}_p}(t') \Bigg\} - \nonumber \\
&& \hspace{0.0cm} - 2 \frac{i}{\hbar} X_{1 \sigma_{{\bf k}_p} {\bf
k}_p,0}(t) \Bigg\{ V_{\text{xx}} \hat{X}_{0,1 \pm {\bf k}}(t) X_{0,1
\pm {\bf k}_p}(t) - i \int_{-\infty}^t dt' F^{\pm \pm}(t-t')
\hat{X}_{0,1 \pm {\bf k}}(t') X_{0,1 \pm {\bf k}_p}(t') \Bigg\}\, ,
\end{eqnarray} where ${\bf k}_i = 2{\bf k}_p-{\bf k}$, and again $V_{\text{xx}}$ and $F^{\pm \pm}(t-t')$ have reabsorbed the $1/2$ originating from Eq.\, (\ref{Sham id per ops}).
In the specific case under analysis we are considering co-circularly
polarized waves and the mean field term, $V_{\text{xx}}$ as well as
the the kernel function $F(t)$ can be found in Refs.\,
\cite{Takayama EPJ, Kwong-Binder PRB 2001}.

Eventually, the lowest order ($\chi^{(3)}$) nonlinear optical
response in SMCs is given by the following set of coupled equations
where, in the same spirit of the final remark in the previous
section, we account for multiple scattering simply by replacing the
linear excitonic polarization and light fields with the total
fields:
\begin{eqnarray}\label{final sys}
&&\frac{d}{dt} \hat{a}_{\pm {\bf k}}  = -i \omega^c_{\bf k}
\hat{a}_{\pm {\bf k}} + i \frac{V}{\hbar}\ \hat{B}_{\pm {\bf k}}
+ t_c \frac{E_{\pm {\bf k}}}{\hbar} \nonumber \\
%==================================================
&&\frac{d}{dt} \hat{B}_{\pm k} = -i\omega_{\bf k} \hat{B}_{\pm {\bf
k}} + \hat{s}_{\pm {\bf k}} + i \frac{V}{\hbar}\ \hat{a}_{\pm {\bf
k}} - \frac{i}{\hbar}{R}^{NL}_{\pm {\bf k}}\, ,\end{eqnarray} where
${R}^{NL}_{\pm {\bf k}}=(R^{sat}_{\pm {\bf k}}+{R}^{\text{xx}}_{\pm
{\bf k}})$
\begin{eqnarray}\label{NN terms}
&& R^{sat}_{\pm {\bf k}} = \frac{V}{n_{\text{sat}}} B_{\pm {\bf
k}_p} a_{\pm {\bf k}_p}
\hat{B}^\dag_{\pm {\bf k}_i} \nonumber \\
&& R^{\text{xx}}_{\pm {\bf k}} = \hat{B}^\dag_{\pm {\bf k}_i}(t)
\bigg( V_{\text{xx}} B_{\pm {\bf k}_p}(t)
B_{\pm {\bf k}_p}(t) - \nonumber \\
&& - i \int_{-\infty}^t dt' F^{\pm \pm}(t-t') B_{\pm {\bf k}_p}(t')
B_{\pm {\bf k}_p}(t') \bigg)\, .
\end{eqnarray} The pump induced
renormalization of the exciton dispersion gives a frequency shift
\begin{eqnarray}
&& \hat{s}_{\pm {\bf k}} = -i \bigg( \frac{V}{n_{\text{sat}}}
\big(B^*_{\pm {\bf k}_p} a_{\pm {\bf k}_p} \hat{B}_{\pm {\bf k}} +
B^*_{\pm {\bf k}_p} B_{\pm {\bf k}_p} \hat{a}_{\pm {\bf
k}} \big) + \nonumber \\
&& \hspace{2.5cm} 2 \frac{V_{\text{xx}}}{\hbar} B^*_{\pm {\bf k}_p}
B_{\pm {\bf k}_p} \hat{B}_{\pm {\bf k}} - \nonumber \\
&& \hspace{2.5cm} -2 \frac{i}{\hbar} B^*_{\pm {\bf k}_p}(t)
\int_{-\infty}^t dt' F^{\pm \pm}(t-t') \hat{B}_{\pm {\bf k}}(t')
B_{\pm {\bf k}_p}(t') \bigg)\, .\end{eqnarray}

Equations, (\ref{final sys}) are the main result of this paper. They
can be considered the starting point for the microscopic description
of quantum optical effects in SMCs. These equations extend the usual
semiclassical  description of Coulomb interaction effects, in terms
of a mean-field term plus a genuine non-instantaneous four-particle
correlation, to quantum optical effects. Analogous equations can be
obtained starting from an effective Hamiltonian describing excitons
as interacting bosons \cite{CiutiBE}. The resulting equations
(usually developed in a polariton basis) do not include correlation
effects beyond Hartree-Fock. Moreover the interaction terms due to
phase space filling differs from those obtaind within the present
approach not based on an effective Hamiltonian.

Only the many-body electronic Hamiltonian, the intracavity-photon
Hamiltonian and the Hamiltonian describing their mutual interaction
have been taken into account. Losses through mirrors, decoherence
and noise due to environment interactions as well as applications of
this theoretical framework will be addressed in another paper of
ours \cite{nostro PRB}.

\section{Conclusion}\label{conclusion}

In this paper we set a dynamics controlled truncation scheme
approach to nonlinear optical processes in cavity embedded
semiconductor QWs without any assumption on the quantum statistics
of the excitons involved. This approach represents the starting
point for the microscopic analysis to quantum optics experiments in
the strong coupling regime. We presented a systematic theory of
Coulomb-induced correlation effects in the nonlinear optical
processes in SMCs. We end up with dynamical equations for exciton
and photon operators which extend the usual semiclassical
description of Coulomb interaction effects, in terms of a mean-field
term plus a genuine non-instantaneous four-particle correlation, to
quantum optical effects. The proper inclusion of the detrimental
environment interactions as well as applications of the present
theoretical scheme  will be presented in another paper of ours
\cite{nostro PRB}.

\appendix
\section{The Equation of Motion At Any Order}\label{gen eq}

The equation of motion for the operators in (\ref{Hubbard}), under
the Hamiltonian $\hat{H} = \hat{H}_{e} + \hat{H}_{c} + \hat{H}_{I} +
\hat{H}_p$ reads:
\begin{eqnarray}
\frac{d}{dt} \left(\right.  \hspace{-0.4cm} &&
\hat{X}_{N\alpha;M\beta} \hat{Y}_{n \lambda;m\mu} \left.\right) =
-i(\omega_{M \beta} -\omega_{N \alpha} + \sum_{i=1}^m \omega^c_{k_i}
- \sum_{j=1}^n \omega^c_{k_j} ) \left(\right.
\hat{X}_{N\alpha;M\beta}
\hat{Y}_{n \lambda;m\mu} \left.\right) + \nonumber \\
&& + \hat{X}_{N\alpha;M\beta} \big( \delta_{m,1} t_c
\frac{E_{\mu}}{\hbar} \hat{Y}_{n \lambda;0} + \delta_{n,1} t_c
\frac{E^{*}_{\lambda}}{\hbar} \hat{Y}_{0;m\mu} \big) -
\hat{X}_{N\alpha;M\beta} \sum_{\bar{k}} t_c \big( \delta_{m,0}
\frac{E^{*}_{\bar{k}}}{\hbar} \hat{Y}_{n \lambda;1
\bar{k}} + \delta_{n,0} \frac{E_{\bar{k}}}{\hbar} \hat{Y}_{1 \bar{k};m\mu} \big) + \nonumber \\
&& + \sum_{\bar{k} \nu} t_c \hat{X}_{N\alpha;M\beta} \Bigg[
\Theta(m-2) \bra{m \mu}\hat{a}^\dag_{\bar{k}} \ket{(m-1)
\nu}\frac{E_{\bar{k}}}{\hbar}
\hat{Y}_{n \lambda;(m-1)\nu} + \nonumber \\
&&\hspace{5.0cm} \Theta(n-2) \bra{(n-1) \nu}\hat{a}_{\bar{k}} \ket{n
\lambda} \frac{E^{*}_{\bar{k}}}{\hbar} \hat{Y}_{(n-1) \nu;m \mu} -
\nonumber \\
&&\hspace{3.0cm} - \Theta(m-1) \bra{m \mu}\hat{a}_{\bar{k}}
\ket{(m+1) \nu} \frac{E^{*}_{\bar{k}}}{\hbar} \hat{Y}_{n \lambda;(m+1)\nu} - \nonumber \\
&& \hspace{5.0cm} \Theta(n-1) \bra{(n+1) \nu}\hat{a}^\dag_{\bar{k}}
\ket{n \lambda} \frac{E_{\bar{k}}}{\hbar} \hat{Y}_{(n+1) \nu;m \mu} \Bigg] + \nonumber \\
&& +\frac{i}{\hbar} \delta_{M,0} \delta_{\beta,0} \delta_{m,1}
\sum_{\bar{n}}V^*_{\bar{n}\mu} \hat{X}_{N\alpha;1 \bar{n} \mu}
\hat{Y}_{n \lambda;0} - \frac{i}{\hbar} \delta_{N,0}
\delta_{\alpha,0} \delta_{n,1} \sum_{\bar{n}} V_{\bar{n}\lambda}
\hat{X}_{1 \bar{n} \lambda;M \beta} \hat{Y}_{0;m \mu} - \nonumber \\
&& -\frac{i}{\hbar} \delta_{N,1} \delta_{n,0}\delta_{\lambda,0}
V^*_{\alpha} \hat{X}_{0;M\beta} \hat{Y}_{1 k_{\alpha};m \mu}
+\frac{i}{\hbar} \delta_{M,1} \delta_{m,0}\delta_{\mu,0} V_{\beta}
\hat{X}_{N \alpha;0} \hat{Y}_{n \lambda;1 k_{\beta}} + \nonumber \\
&& +\frac{i}{\hbar} \delta_{m,1} \Theta(M-1) \sum_{\bar{n} \delta}
V^*_{\bar{n} \mu} \bra{M \beta}\hat{B}_{\bar{n}\mu} \ket{(M+1)
\delta} \hat{X}_{N \alpha;(M+1) \delta} \hat{Y}_{n \lambda,0} -
\nonumber\\
&& -\frac{i}{\hbar}\delta_{n,1} \Theta(N-1) \sum_{\bar{n} \eta}
V_{\bar{n} \lambda} \bra{(N+1) \eta}\hat{B}^\dag_{\bar{n}\lambda}
\ket{N \alpha} \hat{X}_{(N+1) \eta;M \beta} \hat{Y}_{0;m \mu} -
\nonumber\\
&& -\frac{i}{\hbar} \delta_{n,0} \delta_{\lambda,0} \Theta(N-2)
\sum_{\bar{n} \bar{k} \eta} V^*_{\bar{n} \bar{k}} \bra{(N-1)
\eta}\hat{B}_{\bar{n}\bar{k}} \ket{N \alpha} \hat{X}_{(N-1) \eta;M
\beta} \hat{Y}_{1 \bar{k};m \mu} +
\nonumber\\
&& +\frac{i}{\hbar} \delta_{m,0} \delta_{\mu,0} \Theta(M-2)
\sum_{\bar{n} \bar{k} \delta} V_{\bar{n} \bar{k}} \bra{M
\beta}\hat{B}^\dag_{\bar{n}\bar{k}} \ket{(M-1) \delta} \hat{X}_{N
\alpha;(M-1) \delta} \hat{Y}_{n \lambda;1 \bar{k}} +
\nonumber\\
&& +\frac{i}{\hbar} \delta_{M,0} \delta_{\beta,0} \Theta(m-2)
\sum_{\bar{n} \bar{k} \nu} V^*_{\bar{n} \bar{k}} \bra{m
\mu}\hat{a}^\dag_{\bar{k}} \ket{(m-1) \nu} \hat{X}_{N \alpha;1
\bar{n} \bar{k}} \hat{Y}_{n \lambda;(m-1) \nu} -
\nonumber\\
&& -\frac{i}{\hbar} \delta_{N,0} \delta_{\alpha,0} \Theta(n-2)
\sum_{\bar{n} \bar{k} \gamma} V_{\bar{n} \bar{k}} \bra{(n-1)
\gamma}\hat{a}_{\bar{k}} \ket{n \lambda} \hat{X}_{1 \bar{n}
\bar{k};M \beta} \hat{Y}_{(n-1) \gamma;m \mu} -
\nonumber\\
&& -\frac{i}{\hbar} \delta_{N,1} \Theta(n-1) V_{\alpha}
\sum_{\gamma} \bra{(n+1) \gamma}\hat{a}^\dag_{k_{\alpha}}
\ket{n \lambda} \hat{X}_{0;M \beta} \hat{Y}_{(n+1) \gamma;m \mu} + \nonumber\\
&& +\frac{i}{\hbar} \delta_{M,1} \Theta(m-1) V_{\beta} \sum_{\nu}
\bra{m \mu}\hat{a}_{k_{\beta}} \ket{(m+1) \nu} \hat{X}_{N \alpha;0}
\hat{Y}_{n \lambda;(m+1) \nu} + \nonumber\end{eqnarray}
\begin{eqnarray}\label{dt gen op}
&& + \frac{i}{\hbar} \sum_{\bar{n} \bar{k}} \sum_{\nu \delta} {\Bigg
[} V^*_{\bar{n} \bar{k}} {\Big (} \Theta(M-1) \Theta(m-2) \bra{M
\beta}\hat{B}_{\bar{n} \bar{k}} \ket{(M+1) \delta} \nonumber
\\ && \hspace{3.0cm} \bra{m \mu}\hat{a}^\dag_{\bar{k}} \ket{(m-1) \nu} \hat{X}_{N
\alpha;(M+1)\delta} \hat{Y}_{n \lambda;(m-1) \nu} - \nonumber \\
&&  \hspace{2.5cm} - \Theta(N-2) \Theta(n-1) \bra{(N-1)
\delta}\hat{B}_{\bar{n} \bar{k}} \ket{N \alpha} \nonumber \\
&& \hspace{3.0cm} \bra{(n+1)\nu}\hat{a}^\dag_{\bar{k}} \ket{n
\lambda} \hat{X}_{(N-1) \delta;M\beta} \hat{Y}_{(n+1) \nu;m \mu}
{\Big )} + \nonumber \\
&& \hspace{2.0 cm} - \frac{i}{\hbar}V_{\bar{n} \bar{k}} {\Big (}
\Theta(N-1) \Theta(n-2) \bra{(N+1) \delta}\hat{B}^\dag_{\bar{n}
\bar{k}} \ket{N \alpha} \nonumber \\
&& \hspace{3.0cm} \bra{(n-1) \nu}\hat{a}_{\bar{k}} \ket{n \lambda}
\hat{X}_{(N+1) \delta;M \beta} \hat{Y}_{(n-1) \nu;m \mu} - \nonumber \\
&&  \hspace{2.5cm} - \Theta(M-2) \Theta(m-1) \bra{M
\beta}\hat{B}^\dag_{\bar{n}
\bar{k}} \ket{(M-1) \delta} \nonumber \\
&& \hspace{3.0cm} \bra{m\mu}\hat{a}_{\bar{k}} \ket{(m+1) \nu}
\hat{X}_{N \alpha;(M-1)\delta} \hat{Y}_{n \lambda;(m+1) \nu} {\Big
)} {\Bigg ]}\, . \end{eqnarray} Here $\Theta(x)$ is the Heaviside
function equal to 1 for positive argument and zero otherwise.

\section{N {\em eh} pair aggregates}\label{Npair states}

We start from the usual model for the electronic Hamiltonian of a
direct two-band semiconductor \cite{Haugh,AxtKuhn}. It is obtained
from the many-body Hamiltonian of the interacting electron system in
a lattice, keeping explicitly only those terms in the Coulomb
interaction preserving the number of electrons in a given band and
can be expressed as
\begin{equation}\label{Ham electron1} \hat{H}_e = \hat{H}_0 +
\hat{V}_{\text{Coul}}\, .\end{equation}
It comprises the single-particle Hamiltonian terms for electrons in
conduction band and holes in valence band (here $k\equiv ({\bf k},
\sigma)$ and $\hat{c}_{\sigma,{\bf k}}$ ($\hat{d}_{\sigma,{\bf k}}$)
annihilates an electron (a hole)) :
\begin{equation}\label{H zero} \hat{H}_0 = \sum_{k} E_{c,{k}}
\hat{c}^\dag_{k} \hat{c}_{k} + \sum_{k} E_{h,{k}}\hat{d}^\dag_{k}
\hat{d}_{k}\, ,
\end{equation} and the Coulomb interaction term of three
contributions, the two repulsive electron-electron (e-e) and
hole-hole (h-h) terms and the attractive (e-h) one:
\begin{eqnarray}\label{V coul} \hat{V}_{\text{Coul}} && = \frac{1}{2} \sum_{{\bf q}
\neq 0} V_q \sum_{\sigma,{\bf k},\sigma',{\bf k}'}
\hat{c}^\dag_{\sigma,{\bf k}+{\bf q}} \hat{c}^\dag_{\sigma',{\bf
k}'-{\bf q}} \hat{c}_{\sigma',{\bf k}'} \hat{c}_{\sigma,{\bf k}} +
\frac{1}{2} \sum_{{\bf q} \neq 0} V_q \sum_{\sigma,{\bf
k},\sigma',{\bf k}'} \hat{d}^\dag_{\sigma,{\bf k}+{\bf q}}
\hat{d}^\dag_{\sigma',{\bf k}'-{\bf q}} \hat{d}_{\sigma',{\bf k}'}
\hat{d}_{\sigma,{\bf k}} - \nonumber \\
&& - \sum_{{\bf q} \neq 0} V_q \sum_{\sigma,{\bf k},\sigma',{\bf
k}'} \hat{c}^\dag_{\sigma,{\bf k}+{\bf q}}
\hat{d}^\dag_{\sigma',{\bf k}'-{\bf q}} \hat{d}_{\sigma',{\bf
k}'}c_{\sigma,{\bf k}}\, .
\end{eqnarray}
A many-body interacting state is usually very different from a
product state, however a common way to express the former is by a
superposition of uncorrelated product states. The physical picture
that arises out of it expresses the \textit{dressing} the
interaction performs over a set of noninteracting particles. The
general many-body Schr\"{o}dinger equation for this
Coulomb-correlated system is
\begin{equation}\label{Schr eq} \hat{H}_e \ket{\Psi} = (\hat{H}_0 +
\hat{V}_{\text{Coul}}) \ket{\Psi} = E \ket{\Psi}\, ,
\end{equation} with $\ket{\Psi}$ the global interacting many-body
state of the whole Fock space and $E$ its corresponding energy. The
system Hamiltonian commutes with the total-number operators for
electron and holes, i.e. $\hat{N}_e = \sum_{k}
\hat{c}^\dag_{k}\hat{c}_{k}$ and $\hat{N}_h = \sum_{k}
\hat{d}^\dag_{k}\hat{d}_{k}$. Therefore the state $\ket{\Psi}$ may
be build up corresponding on a given number of electrons and of
holes. Moreover, because we shall consider the case of intrinsic
semiconductors materials where $N_e = N_h \doteq N $, the good
quantum number for the Schr\"{o}dinger equation (\ref{Schr eq}) is
the total number of electron-hole pairs $N$, explicitly
\begin{equation}\label{Schr in N} \hat{H}_e \ket{N \alpha} = E_{N
\alpha} \ket{N \alpha}\, , \end{equation} where $\alpha$ is the
whole set of proper quantum numbers needed to specify univocally the
many-body state.

For any given number $N$ of electron-hole pairs, the product-state
set, built up from the single-particle states $\{ \ket{N a} \}$
eigenstates of the noninteracting carrier Hamiltonian $\hat{H}_0$,
is a natural complete basis of the N-pair sector of the global Fock
space:
\begin{equation}\label{nonint H} \hat{H}_0 \ket{N a} =
\epsilon_{N a} \ket{N a}\, , \end{equation} where N identifies the
N-pair subspace and $a$ is a compact form for all the single
particle indexes, i.e. $a \equiv {j}_{e1},{j}_{e2},
...,{j}_{eN};{j}_{h1},{j}_{h2},...,{j}_{hN}$. Indeed
\begin{equation}\label{prod} \ket{N,a} = \otimes_{n=1}^N
\hat{c}^\dag_{j_{en}} \hat{d}^\dag_{j_{hn}} \ket{0,0}\ \ \text{and}
\ \ \epsilon_{N a} = \sum_{n=1}^N (\epsilon_{{j}_{en}} +
\epsilon_{{j}_{hn}} )\, .
\end{equation} Being a complete orthonormal basis for the N-pair
subspace we may expand the many-body state $\ket{N, \alpha}$ over
it, it yields
\begin{equation}\label{U} \ket{N \alpha} = \sum_{a} U^{N \alpha}_{a}
\ket{N a}\, . \end{equation} It is only a matter of calculation to
show that $U^{N \alpha}_a$ is nothing but the envelope function of
the N-pair aggregate, solution of the corresponding secular
equation. Indeed the eigenvalue problem (\ref{Schr in N}) is
transformed into:
\begin{equation}\label{dressing}
\sum_{a'} ( \bra{N a}\hat{H}_e \ket{N a'} - E_{N \alpha}
\delta_{a,a'} ) U^{N \alpha}_{a'} = 0\, . \end{equation} Namely N=1
leads to the exciton secular equation, whereas N=2 represents the
biexciton (two pairs) Coulomb problem.

In order to be clearer we shall propose in details the N=1 exciton
calculation. We shall work in the direct lattice ${\bf r}
\leftrightarrow {\bf r}_i$ (the former is a continuous variable
whereas the latter is a point in the $3D$ lattice). Using the
general mapping \cite{Cohen-Tannoudji QED} $\sum_{r_i}
\leftrightarrow (1/v_0) \int d^3r$, $\delta({\bf r} - {\bf r}') =
(\delta_{{\bf r}_i,{\bf r_j}}/v_0)$, and $ \hat{a}^\dag_{{\bf r}_i}
= (\hat{a}^\dag({{\bf r}_i})/{\sqrt{v_0}}) $, here $v_0$ is the unit
cell volume and for simplicity the spin selection rules for the
optically active states has been already taken into account,
(\ref{dressing}) reads
\begin{equation}\label{sec eq dir l}
\sum_{{\bf r}'_e,{\bf r}'_h} {\Bigg (} \bra{{\bf r}_e,{\bf
r}_h}\hat{H}_e \ket{{\bf r}'_e,{\bf r}'_h} - E_{n \sigma {\bf k}}
\delta_{{\bf r}_e {\bf r}_h,{\bf r}'_e {\bf r}'_h} {\Bigg )}
U^{\alpha}_a({\bf r}'_e,{\bf r}'_h) = 0\, ,
\end{equation} with
\begin{equation}\label{Ham in r} \bra{{\bf r}_e,{\bf r}_h}\hat{H}_e
\ket{{\bf r}'_e,{\bf r}'_h} = {\Bigg (} - \frac{\hbar^2}{2m_e}
\nabla^2_{r_e} - \frac{\hbar^2}{2m_h} \nabla^2_{r_h} -
\frac{e^2}{\varepsilon_r |{\bf r}_e - {\bf r}_h|}+ V({\bf r}_e,{\bf
r}_h) {\Bigg )} \delta_{{\bf r}_e {\bf r}_h,{\bf r}'_e {\bf r}'_h}\,
, \end{equation} here $ V({\bf r}_e,{\bf r}_h)$ represents all the
additional potential, e.g. those of the heterostructures or those of
disorder effects, $ V({\bf r}_e,{\bf r}_h) = V^e(z_e) + V^h(z_h)$.
Typically, the energy difference between the lowest QW subband level
and the first excited one (at least a few meV) is much larger than
the Coulomb interaction between particles (a few meV). As a
consequence, at least at low temperatures, particles are confined at
the lowest quantization level and the (possible) distorsion of the
wave function due to the Coulomb-activated admixture of different
subbands can be safely neglected. In some extent, then, the particle
wave function dependence along the growth (say $z$) direction can be
factorized out and the dynamics becomes essentially two-dimensional.
However, a purely $2D$ approximation for excitons would miss
important effects of the geometrical QWs parameters on the binding
energy and would not be able to account for the interaction with a
$3D$ continuum environment of surrounding modes (e.g. acoustic
phonon modes in heterostructures with alloy lattice constant in
close proximity \cite{Takagahara}). In addition in QWs, light and
heavy holes in valence band are split off in energy. Assuming that
this splitting is much larger than kinetic energies of all the
involved particles and, as well, much larger than the interaction
between them, we shall consider only heavy hole states as occupied.

In Eq.\, (\ref{sec eq dir l}) the $3D$ Coulomb interaction prevents
form factorizing into (free) in-plane and confined directions.
Nevertheless if we assume that the quantization energy along $z$ is
much larger than the Coulomb energy, at leading order we can
factorize out the $z$-dependence
\begin{eqnarray}\label{in z}
{\Bigg (} - \frac{\hbar^2}{2m_e} \frac{d^2}{dz^2_e} + V^e(z_e) -
\frac{\hbar^2}{2m_h} \frac{d^2}{dz^2_h} &+& V^h(z_h) {\Bigg )} U^{\alpha}({\bf r}_e,{\bf r}_h) = \nonumber \\
&& E^{z} U^{\alpha}({\bf r}_e,{\bf r}_h)\, .\end{eqnarray} It means
we are solving our secular equation with solutions built up as
linear combination of $F^{\alpha}_{n_c,n_v,a}({\bf r}^\|_e,{\bf
r}^\|_h)c_{n_c}(z_e)v_{n_v}(z_h)$, with ${\bf r} = ({\bf r}^\|,z)$.
Equation\, (\ref{in z}) expresses the lack of translational symmetry
along the growth $z$ direction, thus single particle states
experience confinement and two additional QW subband quantum numbers
$n_v,n_c$ (for valence and conduction states respectively) appear.
We still leave $a$ as a reminder that new possible indexes could
still arise in due course.

Projecting Eq.\, (\ref{sec eq dir l}) on these confined states we
end up with an effective Schr\"{o}dinger equation in the plane
\begin{eqnarray}\label{Wannier prima}
{\Bigg (} - \frac{\hbar^2}{2m_e} \nabla^2_{\bf r^\|_e} -
\frac{\hbar^2}{2m_h} \nabla^2_{\bf r^\|_h} -
U_{n_c,n_v;n'_c,n'_v}(|{\bf r}^\|_e - {\bf r}^\|_h|) {\Bigg )}
F^{\alpha}_{n_c,n_v,a}({\bf r}^\|_e,{\bf
r}^\|_h) = \nonumber \\
&& \hspace{-5.0cm} = (E_{\alpha} - E^z_{n_c} - E^z_{n_v})
F^{\alpha}_{n_c,n_v,a}({\bf r}^\|_e,{\bf r}^\|_h)\, ,\end{eqnarray}
with \begin{equation} U_{n_c,n_v;n'_c,n'_v}(|{\bf r^\|_e} - {\bf
r^\|_h|}) = \int dz_e \int dz_h \frac{e^2}{\varepsilon_r \sqrt{|{\bf
r}^\|_e - {\bf r}^\|_h|^2 +(z_e - z_h)^2}} c_{n_c}(z_e)
c_{n'_c}(z_e) v_{n_v}(z_h)v_{n'_v}(z_h)\, .
\end{equation}
For what already stated, we shall consider only the lowest confined
subband levels, then the resulting effective in-plane secular
equation becomes
\begin{eqnarray}\label{Wannier GS}
{\Bigg (} - \frac{\hbar^2}{2m_e} \nabla^2_{\bf r^\|_e} -
\frac{\hbar^2}{2m_h} \nabla^2_{\bf r^\|_h} - \int dz_e \int dz_h
\frac{e^2}{\varepsilon_r \sqrt{|{\bf r}^\|_e - {\bf r}^\|_h|^2 +(z_e
- z_h)^2}} |c_{n_c}(z_e)|^2 |v_{n_v}(z_h)|^2 {\Bigg )}
F^{\alpha}_{a}({\bf r}^\|_e,{\bf r}^\|_h) = \nonumber \\
&& \hspace{-8.0cm} = (E_{\alpha} - E^z_c - E^z_v)
F^{\alpha}_{a}({\bf r}^\|_e,{\bf r}^\|_h)\, ,\end{eqnarray}with the
product exciton envelope function $U^{\alpha}({\bf r}_e,{\bf
r}_h)=F^{\alpha}_{a}({\bf r}^\|_e,{\bf r}^\|_h)c(z_e)v(z_h)$.
Equation\, (\ref{Wannier GS}) is solvable by separation of variables
once we employ a coordinate transformation into center of mass (CM)
${\bf R}=(m_e{\bf r}^\|_e+m_h{\bf r}^\|_h)/(m_e+m_h)$ and relative
${\bf \rho}=({\bf r}^\|_e - {\bf r}^\|_h)$ exciton coordinates. It
reads
\begin{equation}\label{Wannier R rho}
{\Bigg (} - \frac{\hbar^2}{2M} \nabla^2_{\bf R} -
\frac{\hbar^2}{2\mu} \nabla^2_{\bf \rho} - U({\bf \rho}) {\Bigg )}
F^{\alpha}_{a}({\bf R}, {\bf \rho}) = E F^{\alpha}_{a}({\bf R}, {\bf
\rho})\, ,\end{equation} with a solution we can arrange as
$F^{\alpha}_{a}({\bf R}, {\bf \rho}) = \frac{e^{i\, {\bf K} \cdot
{\bf R}}}{\sqrt{A}} W^{\alpha}_a({\bf \rho})$ the latter solution of
the relative hydrogen-like 2D problem.

Eventually, in real-space representation, we have our exciton wave
function with total in-plane CM wave vector ${\bf K}$ ($A$ is the
in-plane quantization surface in the free directions) which reads
\begin{equation}\label{exciton} \ket{n \sigma {\bf K}}= \frac{v_0}{\sqrt{A}}
\sum_{\bf r_e,r_h}  e^{i \bf k \cdot R} W_{n \sigma}(\rho) c(z_e)
v(z_h) a^\dag _{c,{\bf r}_e} a^\dag_{h,{\bf r}_h } \ket{0}\, ,
\end{equation} being $\hat{a}^\dag _{c/v,{\bf r}}
(\hat{a}_{c/v,{\bf r }})$ creation (annihilation) operator of the
conduction- or valence- band electron in the Wannier representation
and ${\bf r}_{e/h} = ({\bf r}^{\|}_{e/h},z_{e/h})$ are to be
considered coordinates of the direct lattice, $\ket{\!0}$ is the
crystal ground state.

When e.g. exploring the exciton-phonon interaction, it is useful to
express exciton states in reciprocal space. With the usual
transformation to Bloch representation, ($N = v_0 A\,L$ is the
number of unit cells and $L$ is the quantization dimension along the
confined direction, $\nu = c,v$),
\begin{equation}\label{BlochWannier} \hat{a}_{\nu ({\bf r},z)} = \frac{1}{\sqrt{N}} \sum_{{\bf k},k_z} e^{i
{\bf k} \cdot {\bf r}_n} \hat{a}_{\nu ({\bf k},k_z)}\, ,
\end{equation} one obtains:
\begin{eqnarray}\label{exc in Bloch rep 1}
&& \ket{n \sigma {\bf K}} = \sum_{\substack{{\bf k}, {\bf
k}'\\{k_z,k'_z}}} \delta_{{\bf K},{\bf k}-{\bf k}'} {\Bigg (}
\frac{1}{\sqrt{A}\, L} \int d{\rho} \int dz_e \int dz_h W_{n
\sigma}(\rho) c(z_e) v(z_h) e^{-i\, {\bf \rho} \cdot (\eta_h {\bf k}
+ \eta_e {\bf k}')} e^{-i\, k_z z_e} e^{-i\, k'_z z_h} {\Bigg )}
\nonumber \\
&& \hspace{3.0cm} a^\dag _{c,({\bf k},k_z)}a_{v,({\bf k}',k'_z)}
\bigl| 0\bigr>\, .\end{eqnarray} In order to end up with a form as
much as possible in analogy with its bulk counterpart we shall
define CM and relative coordinates even in the reciprocal lattice:
\begin{eqnarray}\label{rules}
\left\{ \begin{array}{l} {\bf K} = {\bf k} - {\bf k}' \\
{\bf k}_r = \eta_h {\bf k} + \eta_e {\bf k}' \end{array} \right.
\Longrightarrow \left\{ \begin{array}{l} {\bf k} = {\bf k}_r + \eta_e {\bf K} \\
{\bf k}' = {\bf k}_r - \eta_h {\bf K} \end{array} \right.\, .
\end{eqnarray}
It becomes
\begin{eqnarray}\label{exc in Bloch rep 2}
&& \ket{n \sigma {\bf K}} = \sum_{{\bf K}, {\bf k}_r} \delta_{{\bf
K},{\bf k}-{\bf k}'} \sum_{k_z,k'_z} {\Bigg (} \frac{1}{\sqrt{A}}
\int d{\rho} W_{n \sigma}(\rho) e^{-i\, {\bf \rho} \cdot {\bf k}_r}
{\Bigg )} {\Bigg (} \frac{1}{\sqrt{L}} \int dz_e c(z_e) e^{-i\, k_z
z_e} {\Bigg )} \nonumber \\
&& \hspace{3.0cm} {\Bigg (} \frac{1}{\sqrt{L}}\int dz_h v(z_h)
e^{-i\, k'_z z_h} {\Bigg )} a^\dag _{c,({\bf k}_r + \eta_e {\bf
K},k_z)}a_{v,({\bf k}_r - \eta_h {\bf K},k'_z)} \bigl| 0\bigr>\,
.\end{eqnarray}
Thus
\begin{eqnarray}\label{exc in Bloch rep final}
&& \ket{n \sigma {\bf K}} = \sum_{{\bf k}_r} \sum_{k_z,k'_z}
\Phi^{\bf K}_{n \sigma,{\bf k}_r} u^c_{k_z} u^{v *}_{k'_z} a^\dag
_{c,({\bf k}_r + \eta_e {\bf K},k_z)}a_{v,({\bf k}_r - \eta_h {\bf
K},k'_z)} \bigl| 0\bigr>\, ,\end{eqnarray} or in the electron-hole
picture ($\hat{a}_{v,{\bf k}} = \hat{d}^\dag_{-{\bf k}}$ and $-{\bf
k}\left.\right|_{el} = {\bf k}\left.\right|_{hole}$)
\begin{eqnarray}\label{exc in Bloch rep final hole picture}
&& \ket{n \sigma {\bf K}} = \sum_{{\bf k}_r} \sum_{k_z,k'_z}
\Phi^{\bf K}_{n \sigma,{\bf k}_r} u^e_{k_z} u^{h}_{k'_z} c^\dag
_{({\bf k}_r + \eta_e {\bf K},k_z)}d^\dag_{(-{\bf k}_r + \eta_h {\bf
K},-k'_z)} \bigl| 0\bigr>\, ,\end{eqnarray} with the relations in
Eq.\, (\ref{rules}) changed accordingly.


\begin{thebibliography}{999}
\bibitem{detail} Inserting two photon identities at both sides
$\hat{a}_{k} = \sum_{n,\gamma}
Y_{n,\gamma;n,\gamma}\hat{a}_{k}\sum_{m,\lambda}
Y_{m,\lambda;m,\lambda}$, where $n= \sum_{i= modes} n_i$ and
$m=\sum_{i =modes} m_i$, we have non-zero matrix elements only if
$n_i = m_i \forall i \neq k$ and $n_k + 1= m_k$. Thus it becomes
$\hat{a}_{k} = \sum_{(n_k,n'),\gamma} (\sqrt{{n_k}+1} Y_{n_k
k;(n_k+1) k}) \otimes Y_{n' \gamma; n' \gamma}$, where $n'$ stands
for the string $(n_1,n_2, ...)$ without the $k$-th entry. When we
make  the one photon addendum explicit, we end up with $Y_{0;1 k}$.
\end{thebibliography}
\end{document}